\documentclass[12pt,preprint]{aastex}

\begin{document}

\title{Cloud Structure and Physical Conditions in Star-forming Regions
       from Optical Observations. ~II. Analysis}

\author{K. Pan\altaffilmark{1,2}, 
S.R. Federman\altaffilmark{1},
Y. Sheffer\altaffilmark{1}, 
and B-G Andersson\altaffilmark{3}}
\altaffiltext{1}{Department of Physics and Astronomy, University of
Toledo, Toledo, OH 43606; steven.federman@utoledo.edu; 
ysheffer@physics.utoledo.edu.}
\altaffiltext{2}{Current address: Department of Physics and Astronomy,
Bowling Green State University, Bowling Green, OH 43403; 
kpan@hubble.bgsu.edu.} 

\altaffiltext{3}{Department of Physics and Astronomy, Johns 
Hopkins University, Baltimore, MD 21218; bg@pha.jhu.edu.}

\setcounter{footnote}{3}

\begin{abstract}
To complement the optical absorption-line survey of diffuse molecular gas in
Paper I, we obtained and analyzed far ultraviolet H$_2$ and CO data on lines 
of sight toward stars in Cep OB2 and Cep OB3. Possible correlations between 
column densities of different species for individual velocity components, 
not total columns along a line of sight as in the past, were examined 
and were interpreted in terms of cloud structure. The analysis reveals 
that there are two kinds of CH in diffuse molecular gas: CN-like CH and 
CH$^+$-like CH.  Evidence is provided that CO is also associated with CN 
in diffuse molecular clouds. Different species are distributed according
to gas density in the diffuse molecular gas. Both calcium and potassium may 
be depleted onto grains in high density gas, but with different dependences 
on local gas density. Gas densities for components where CN was detected 
were inferred from a chemical model. Analysis of cloud structure indicates 
that our data are generally consistent with the large-scale structure 
suggested by maps of CO millimeter-wave emission. On small scales, the gas 
density is seen to vary by factors greater than 5.0 over scales of 
$\sim$ 10,000 AU. The relationships between column densities of CO and 
CH with that of H$_2$ along a line of sight show similar slopes for the 
gas toward Cep OB2 and OB3, but the CO/H$_2$ and CH/H$_2$ ratios tend 
to differ which we ascribe to variation in average density along the 
line of sight.

\end{abstract}


\section{INTRODUCTION}

The physical and chemical state in the interstellar medium (ISM) 
continuously changes through the combination of star 
formation and stellar death (Cox \& Smith 1974; Jenkins \& Meloy 1974;
McKee 1990; Thornton et al. 1998; Marri \& White 2004). The evolution of 
a galaxy is governed to a large extent by 
these processes. The final stages of star formation involve the disruption 
of interstellar material from which the stars formed. If enough material is 
available in interstellar clouds forming O and B stars, one generation of 
stars can lead to the formation of a second generation, and so forth, but 
eventually the cloud remnants can no longer sustain star
formation. The propagation of star formation occurs in regions of enhanced 
gas density behind shocks (e.g. Elmegreen \& Lada 1977; McKee \& Tan 2003 ). 
Early type  (O and B) stars are associated with stellar winds, 
expanding \ion{H}{2} regions, and supernova explosions that lead to shocks 
in the surrounding medium. Since these stars tend to form in 
clusters (Clark et al. 2005), the effects are magnified 
as seen in the Magellanic Clouds. 
Clearly, knowledge of the physical conditions and 
chemical composition of the ISM in star-forming regions will help us 
understand the above processes and their effects on the evolution of the 
ISM and the Galaxy. 

High-resolution optical observations of interstellar absorption (e.g.,
Welty \& Hobbs 2001; Pan et al. 2004, hereafter Paper I) have revealed
complex velocity structure on many Galactic lines of sight, especially
on sight lines through star-forming clouds. 
Determination of the physical properties of 
individual interstellar clouds from observations of absorption lines 
requires high spectral resolution
to distinguish the individual components contributing to the 
generally complex line profiles. Furtheremore, ultraviolet (UV) spectra 
provide considerable information on abundances and general
physical conditions in the ISM (e.g., Spitzer \& Jenkins 1975). In 
Paper I, we presented high-resolution optical spectra of interstellar 
CN, CH, CH$^{+}$, 
\ion{Ca}{1}, \ion{K}{1}, and \ion{Ca}{2} absorption along 29 lines of
sight in three star-forming regions, $\rho$ Oph, Cep OB2, and Cep OB3.
We obtained velocity component structure by simultaneously 
analyzing the spectra of the six species 
along a given sightline. Column densities and Doppler parameters of the 
individual velocity components were derived through profile fitting. To 
complement the results of the optical survey, presented in Paper I, we here 
present an analysis of far ultraviolet (FUV) spectra of 15 of the stars in that 
sample.  The FUV data were either acquired in dedicated observing programs or 
extracted from the existing data archives of the {\it Far Ultraviolet 
Spectroscopic Explorer} (FUSE) and {\it Hubble Space Telescope/Space 
Telescope Imaging Spectrograph} (HST/STIS).  
The main contribution of the present paper is an
analysis of the results from Paper I and those for H$_2$ and CO in terms
of cloud structure for gas with densities of 10 to 1500 cm$ ^{-3}$.
We emphasize that our data for the diffuse molecular
gas associated with the three star-forming regions probe the large-scale
structure of the parent molecular cloud as it disperses. We investigate
whether a connection exists between the levels of star formation and
conditions in the surrounding diffuse gas. 

The remainder of the paper has the following outline. The next section 
gives a brief background of the three star-forming regions, 
and $\S$ 3 describes the FUV observations and
their analysis. A comparison of Doppler parameters is the focus of
$\S$ 4. In $\S$ 5, we examine correlations between column 
densities of different species. A chemical analysis is performed in $\S$ 6
to derive gas densities for components with CH and CN.  
The analysis utilizes available results for C$_2$ from the literature.
The amount of H$_2$ excitation and its relationship to the chemical
results is described here also. This
is followed by a discussion on the distribution of species within clouds, 
on large- and small-scale cloud structures, and a comparison of general 
properties for the molecular gas in 
the three star-forming regions. The final section presents a summary of 
our work.   

\section {Background}

The divergence in the modes of star formation is
perhaps best understood in terms of the physical properties of
the molecular cloud and environmental effects such as nearby \ion{H}{2}
regions. $\rho$ Oph, Cep OB2, and Cep OB3 are three 
regions that provide fruitful laboratories for the investigation of 
cloud dispersal as a result of star formation. Our data on diffuse 
molecular gas, whose $LSR$ velocities are similar to those of
the radio-emitting clouds, probe the later stages of dispersal. 
Extensive studies (e.g., 
Sargent 1979; de Geus, Bronfman, \& Thaddeus 1990; Patel et al. 1998) 
on these regions outlined basic properties for the cloud complexes.  

\subsection {The $\rho$ Oph Region}

The region around the star $\rho$ Oph, also called the $\rho$ Oph Molecular 
Cloud, is a well-known concentration of dark nebulae and molecular clouds
with a size of about 6$\times$5 pc$^2$, a total molecular mass of 80
$M_{\odot}$ (de Geus et al. 1990), and at a distance of 140 pc. One of 
its prominent characteristics 
is the extremely active star formation in the cloud core, where stars
are 
heavily clustered with an estimated star-forming efficiency 
$\ge$ 20\% (cf. Green \& Young 1992; Doppmann et al. 2003; Phelps 
\& Barsony 2004). Radio 
observations (Loren 1989; de Geus et al. 1990) showed that CO molecules 
mainly exist within a velocity range of $V_{LSR} \sim$ 1.5--5.0 km s$^{-1}$, 
and that the cloud  forms the boundary between the bulk of the Ophiuchus 
molecular gas and the Upper-Scorpius stellar groups.
 
The $\rho$ Oph Molecular Cloud is part of the Ophiuchus molecular clouds, 
a filamentary system of clouds with a total molecular mass of 
10$^4$ $M_{\odot}$ (de Geus 1992). The Ophiuchus molecular clouds, 
in turn, belong to a larger cloud complex that 
is associated with the Scorpio-Centaurus OB association. This OB association 
consists of three subgroups, Lower-Centaurus Crux (LCC), 
Upper-Centaurus Lupus (UCL), and Upper-Scorpius (US). Blaauw (1958)
showed that the three subgroups are separated in position and age, and that 
the youngest subgroup lies in Upper-Scorpius, which is adjacent to the 
$\rho$ Oph Molecular Cloud. Based on photometric measurements, de Geus, 
de Zeeuw, \& Lub (1989) derived ages of 11--12, 14--15, and 5--6 Myr for the 
three subgroups, respectively. Somewhat surprisingly, they found that the 
middle subgroup, UCL, is the oldest one. They argued that either the 
classic picture of sequential star formation (Blaauw 1964; Elmegreen \&
Lada 1977) is not valid for the Sco OB2 Association as a whole because the 
oldest subgroup is inbetween the two younger ones, or for some unknown 
reason, massive star formation was initiated near the middle of the 
original giant molecular cloud. However, Pan (2002) conjectured that 
LCC and UCL may actually have a similar age.  
More recently, Sartori, Lepine, \& Dias (2003) found that the US 
subgroup is 4\sbond 8 Myr old and that UCL and 
LCC have the same age, 16-20 Myr. 
 
\subsection{The Cepheus Bubble}

A region of enhanced infrared emission associated with Cep OB2
shows the location of a giant ring-shaped 
cloud system, the so called Cepheus Bubble. The distance of the Bubble is 
somewhat uncertain. Earlier measurements indicated distances from 700 to 
900 pc (Garrison \& Kormendy 1976; Kun, Balazs, \& Toth 1987). Recently,
de Zeeuw et al. (1999) determined the distances of nearby OB associations 
using proper motion and parallaxes measured by Hipparcos. They 
obtained a distance of 615 pc for Cep OB2. As de Zeeuw et al. noticed,
their distances are 
systematically smaller than the previous photometric determinations.  
If a distance of 750 pc is adopted, the ring-shaped cloud system has a 
diameter of about 120 pc and the total molecular mass for the system is about 
10$^5$ $M_{\odot}$ (Patel et al. 1998)\footnote{If the distance of 615 pc, 
as suggested by de Zeeuw et al. (1999), is adopted, the size and mass of 
the cloud system will be slightly smaller. Our discussion on large- and 
small-scale structure will not be affected by the choice of the distance.}. 
A dozen star-forming regions have been identified in the
rim of the Bubble. Each star-forming region has a size of a few pc, similar 
to that of  the $\rho$ Oph region, and total molecular mass of  about 
10$^3$ $M_{\odot}$. Some large star-forming regions consist of several 
star-forming globules. For instance, IC 1396 includes at least four 
star-forming globules with a total mass of about 2200 $M_{\odot}$ and 
a size of 12 pc (Patel et al. 1995); Weikard et al. (1996) reached
similar conclusions. 

Based on measurements of CO emission, Patel et al. (1998) proposed that 
the Bubble was created by stellar winds, and likely supernova explosions, 
from now deceased stars in the cluster NGC 7160 as well as the evolved 
stars, VV Cep and $\mu$ Cep, in the region. Patel et al. (1998) suggested 
that these stars were the first generation formed. The first generation 
stars have an age of about 13-18 Myr. After about $\sim$ 7 Myr, the 
expanding shell reached a size of about 30 pc and became unstable. The 
instability led to the formation of a second generation of stars, current 
numbers of Cep OB2, which are about $\sim$ 7 Myr old. The IRAS point 
sources in the globules and molecular clouds associated with the Bubble 
are the third generation, whose formation was triggered by stellar winds 
from Cep OB2 stars. From \ion{H}{1} 21 cm data, Abraham, Balazs 
\& Kun (2000) obtained 
a similar picture for the formation of the Bubble. The latter model 
showed that a supernova explosion might have occurred as recently as 
about 2 Myr ago. The explosion expanded the pre-existing Bubble further.

\subsection{Cloud System Associated with Cep OB3}

The molecular clouds producing the Cep OB3 Association represent a 
filamentary system, showing clumpy and irregular structure 
(Sargent 1977, 1979). The cloud system is located at a 
distance of about 700--800 pc (Garrison 1970; Moreno-Corral et al. 1993), 
with a size of $\sim$ 
60$\times$20 pc$^2$ and a total molecular mass of 
$\sim$ 10$^4$ $M_{\odot}$ (Sargent 1979). A few identified star-forming 
regions are embedded in the cloud 
system. The star-forming regions in this cloud system have a size similar
to those in the Cepheus Bubble (a few pc), but have smaller total 
molecular mass (100--500 $M_{\odot}$). Our lines of sight toward Cep 
OB3 probe two star-forming regions in the cloud system, Cepheus B and 
Cepheus F, with total molecular mass of 100 and 300,
respectively (Yu et al. 1996). 

The Cep OB3 Association consists of two subgroups separated by $\sim$ 
13 pc on the plane of the sky. Based on photometric measurements, 
Blaauw (1964) assigned ages of 8 and 4 Myr to the two subgroups. However, 
ages derived from such isochrone fitting must be treated with caution in 
young OB associations
because masses and ages remain model dependent. 
Consequently, the shape of both the evolutionary tracks and of the 
isochrones change from one model to another, leading to discrepancies 
in the age estimates. Much lower ages for the younger subgroup can be 
found in the literature. For instance, Garmany (1973) and Assousa,
Herbst \& Turner (1977) obtained an age of $\sim$ 1 Myr for it. 
Sargent (1979) suggested 
that the formation of the younger group was triggered by the interaction 
between stellar winds of the older subgroup and the original molecular 
cloud.

\section{FUV Observations and Analysis}

In Paper I, we presented observational data for interstellar 
CN, CH, CH$^{+}$, \ion{Ca}{1}, \ion{K}{1}, and \ion{Ca}{2} absorption on 
29 directions in $\rho$ Oph, Cep OB2, and Cep OB3. To 
complement these results, 
we obtained and analyzed FUV data for H$_2$ and CO. 
Eight stars from the sample in Paper I were observed
with {\it FUSE} under our programs A051 and B030.
For each star, reduction and calibration were performed using CALFUSE V2.4.
We then re-binned the data by a factor of 4, yielding a two-pixel 
resolution element of $\sim$ 0.06 \AA.
{\it FUSE} archival data for two more stars, HD~206773 and HD~217312, were
downloaded from programs B071 and P193, respectively.
Four additional stars have been studied before; we used their
published H$_2$ column density, $N$(H$_2$), and $T_{1,0}$ results from
Rachford et al. (2003) (under {\it FUSE} program P116).
Although no published H$_2$ results are available for HD~208266, we included 
this star in our analysis because there were CH and CO observations.
Moreover, these data can be utilized in tandem to predict $N$(H$_2$) 
toward HD~208266. Table 1 lists all FUV datasets that we utilized, 
specifying which space telescope was the source, what S/N was obtained, 
and which molecules were analyzed.

Three $B-X$ bands of H$_2$ (2$-$0, 3$-$0, and 4$-$0, between 1042 and 1083 
\AA) were chosen to model the molecule's column density.
This range is covered by the four overlapping {\it FUSE} detector segments,
LiF-1A, LiF-2B, SiC-1A, and SiC-2B, although 2$-$0 is not covered by 
LiF-2B. Each segment was fitted independently with the code ISMOD, which 
uses the simplex method to minimize the rms of the residuals down to about 
10$^{-4}$ in relative parameter steps. The fit was based on the CH cloud 
components in Paper I, preserving the velocity
separations and relative fractions in column density among components
as fixed input. We, however, allowed the $b$-values 
to vary during the fits, constraining them by the mass ratio of CH to 
H$_2$ and by the kinetic temperature ($T$) in the gas as given by the 
(fitted) $T_{1,0}$ rotational temperature of H$_2$ (see $\S$ 6.2).
Both the thermal and non-thermal components of the velocity field were 
thus determined. Free parameters included the total column density, 
$N$(H$_2$), radial velocity
of H$_2$, six rotational temperatures relative to the ground state,
$T_{J^\prime}$$_{^\prime,0}$, and the placement of the stellar 
continuum. Our final H$_2$ model and the associated uncertainties 
were computed by averaging the results from fits of the 
four independent segments. In a few cases of high-J levels, where one 
of the four segment yielded a deviant result, that value was not included 
in the final average. The values of total $N$(H$_2$) are very 
robust and practically independent of the cloud structure because of the 
damping wings for the $J^\prime$$^\prime$ = 0 and 1 lines.

For determinations of CO column density, we first searched the 
{\it HST}/STIS
archive and downloaded high-resolution observations of $A-X$ bands of CO.
Six stars happen to have E140H data with R $\ga$ 100,000, covering $A-X$ 
bands blueward of $\sim$ 1360 \AA\ (transitions 7$-$0 and higher).
For the ISMOD fits, we used $f$-values from Chan, Cooper \& Brion (1993), 
see also Morton and Noreau (1994). These values are within a few percent 
of those recommended by Eidelsberg et al. (1999), except for the
$f$-value for 11$-$0 where Chan et al.'s value is 11\% smaller.  
The higher {\it HST}/STIS 
resolution allowed us to derive relative cloud strengths as well. 
When the CO along a line of sight 
is relatively weak, not enough fitting
leverage is available from the $A-X$ bands alone.
In such cases we expanded the analysis to include simultaneous fits of Rydberg 
bands from the {\it FUSE} data with R $\approx$ 17,500.
For CO toward HD~203374A, our simultaneous fits of $A-X$ and Rydberg bands 
were published in Sheffer, Federman, \& Andersson (2003). An 
additional three stars have medium-resolution data from the 
E140M grating (R = 46,000).
Despite the lower resolution, robust fits are reliably obtainable in these
cases because the spectral coverage includes 14 $A-X$ bands
down to the lowest S/N region around Ly-$\alpha$.
Nevertheless, in our fits we employed at most six bands for simultaneous
solutions, making sure to include bands spanning the largest possible range 
in $f$-value. Adding more bands to the fit did not change the results 
significantly. Finally, for the six stars without any archival 
{\it HST}/STIS data, our CO modeling was based solely on the three 
Rydberg bands, $B-X$ 0$-$0, $C-X$ 0$-$0, and $E-X$ 0$-$0.
Understandably, the inferred CO component structure for these stars are 
less reliable than for those from $A-X$ bands, due to the 
lower resolution of {\it FUSE} and the limited range in $f$-values sampled, 
which were based on the results of Federman et al. (2001).
However, since the $B-X$ 0$-$0 band is relatively optically thin, 
the total $N$(CO) is approximately independent of the cloud structure.
The derived total H$_2$ and CO column densities along the fifteen lines of
sight are listed in Table 2, along with the kinetic temperature ($T_{1,0}$)
of H$_2$. For comparison, we also tabulate CH column densities 
from Paper I and abundance
ratios (CH/H$_2$ and CO/H$_2$) along these sight lines. Table 3 provides 
the column densities for individual H$_2$ rotational levels.

\section{Doppler Parameter and CN Excitation Temperature}

The Doppler parameters, $b$-values,  of individual absorption components 
seen for any individual species provide upper limits on the temperature 
and internal turbulent velocity ($v_t$) in the interstellar gas because 
$b = (2kT/m + v_t^2)^{1/2}$. If two species with significantly different
atomic weight $m$ coexist in the same volume of gas, comparison of their
$b$-values provides estimates of the relative contributions of thermal
and turbulent broadening, if both lines are fully resolved. Alternately, 
known temperatures and turbulent
velocities for two species can allow estimates of their relative volumetric 
distributions. The extracted $b$-values in the
present paper, except for CN $b$-values from the doublet ratio method in this 
section, are based on Gaussian instrumental widths.

Earlier studies suggested that CN resides in
denser regions whereas CH$^{+}$ is found mainly in regions of lower density
(Cardelli et al. 1990; Federman et al. 1994). The CH molecule, on the 
other hand, can be present in both low- and high-density gas (100 vs.
600 cm$^{-3}$). It can be 
synthesized through the non-equilibrium CH$^{+}$ chemistry in low density 
diffuse clouds (Draine \& Katz 1986; Zsarg\'{o} \& Federman 2003), and in 
moderately 
dense gas it is produced via C$^{+}$ + H$_{2} \rightarrow$ CH$^{+}_{2} + 
h\nu$ (Federman et al. 1994).  However, it is not a trivial task to 
disentangle the amount of CH formed with CH$^{+}$ or associated with CN. 
Examining CH$^{+}$--like CH and CN--like CH components may elucidate the 
chemical schemes occurring in the gas (Lambert, Sheffer, \& Crane 1990). 
Of our 125 CH components, there were 12 CN--like CH components, 
which are CH components with corresponding CN, but 
no CH$^{+}$ [$N$(CH$^+$) $ < 3.0 \times 10^{11}$ cm$^{-2}$], at the same
$V_{LSR}$, and 78 CH$^{+}$--like CH components, CH components with detected
CH$^{+}$ at the same V$_{LSR}$ but no CN [$N$(CN) $ < 2.0 \times 10^{11}$ 
cm$^{-2}$]. Typically, CH$^+$ and CN column densities are 7
times greater than these limits. Table 4 presents average 
$b$-values  of the CH$^{+}$-like CH and the CN-like CH components, along 
with mean Doppler parameters of all defined components for other species.
The analysis reveals that CN-like CH components have an average $b$-value 
indistinguishable from that of CN components (0.83 $\pm$ 0.11 km s$^{-1}$ 
vs. 0.90 $\pm$ 0.11 km s$^{-1}$), whereas CH$^+$-like CH components 
possess a larger average $b$-value (1.10 $\pm$ 0.16 km s$^{-1}$). 
Simple statistical tests show that the average $b$-values 
for a given species agree within their mutual uncertainties in the diffuse gas 
surronding different 
star-forming regions, although Cep OB3 seems to have consistently larger 
$b$-values than $\rho$ Oph and Cep OB2. Since thermal broadening plays a 
minor role, our results indicate that cloud turbulence in the three 
regions does not differ significantly. 

Compared with earlier studies, our average $b$-values of
atomic species (Table 4) are slightly greater than those obtained from
very high-resolution spectra of interstellar atomic lines. 
Based on spectra with resolution of 0.3-0.6 km s$^{-1}$, Welty and
colleagues obtained median $b$-values of 0.66 km s$^{-1}$ for
\ion{Ca}{1} components (Welty, Hobbs, \& Morton 2003), 0.67 km s$^{-1}$ 
for \ion{K}{1} (Welty \& Hobbs 2001),  and 1.33 km s$^{-1}$
for \ion{Ca}{2} components  (Welty, Morton, \& Hobbs 1996). Three
causes may lead to the differences. Their spectra had much 
higher resolution than ours so that they could discern more closely blended 
components which we did not, different lines of sight were probed, and the
median values are slightly smaller than corresponding average values. 
(For example, the median $b$ is 0.66 km s$^{-1}$ while 
the corresponding average is 0.70 km s$^{-1}$ for well defined \ion{Ca}{1} 
components.)  With a resolution of $\sim$ 0.50 km s$^{-1}$, Andersson, 
Wannier, \& Crawford (2002) obtained spectra of interstellar
CH absorption toward 18 stars in southern molecular cloud envelopes.
Based on F-tests (Lupton 1993), they fitted all but one
of these CH profiles with a single-velocity component. (However, they
noticed that additional velocity components may be present). Their 
$b$-values range from 0.3 to 3.6 km s$^{-1}$, with an average of 1.7 
km s$^{-1}$. This mean $b$-value is greater than ours, 1.04 km s$^{-1}$ 
(Table 4). Because we simultaneously analyzed the spectra of six species 
along a given sightline (Paper I), in general, we found more CH velocity
components per line of sight. This is one of the reasons, if not main one,
that our average $b$ value is smaller than theirs. 
Based on ultra-high-resolution ($\delta$v $\sim$ 0.35 km s$^{-1}$) 
observations along five lines of sight, Crawford (1995) found average 
$b$-values of 0.63, 1.5, 2.3, and 0.8 km s$^{-1}$ for CN, CH, CH$^+$, and 
\ion{K}{1} components. Overall, our average Doppler parameters are in 
agreement with those determined by previous studies using spectra with
higher resolution.

  The $b$-value of a CN component can also be derived by using the doublet 
ratio (Str\"{o}mgren 1948) --- see also Gredel, van Dishoeck, \& Black (1991). 
Because CN $R$(1) and $P$(1) 
absorption lines arise from the same rotational level, $N^{\prime\prime}$=1, 
the same column density should be inferred from them. Therefore, the $b$-value 
of a CN component can be determined by performing a curve of growth analysis 
on the $R$(1) and $P$(1) values of $W_{\lambda}$ and requiring they give 
the same 
column density and $b$-value. To check $b$-values of CN components derived 
from the profile fitting in Paper I, we applied the doublet ratio
method to eleven components where both CN $R$(1) and $P$(1) lines 
were detected and 
obtained an average $b$-value of 0.70 $\pm$ 0.14 km s$^{-1}$, which is 
closer to that of Crawford (1995) based on ultra-high-resolution spectra.
Considering the uncertainties, it also agrees with the mean Doppler parameter 
in Table 4, which we determined from profile fitting.   

Once the $b$-value of a CN component is determined, column densities of 
rotational 
levels of $N^{\prime\prime}$=0 [from CN $R$(0)] and $N^{\prime\prime}$=1 [from 
CN $R$(1) and $P$(1)] can be derived by using a curve of growth. 
Knowing columns of these rotational levels 
allows us to obtain the CN excitation temperature, $T_{ex}$, of
CN components. $T_{ex}$ is governed by the Boltzmann equation,

\begin{equation}
 T_{ex} = \frac{h\nu}{k_{B}}\left(ln\left[\frac{g_{1}N(0)}{g_{0}N(1)}\right]
 \right)^{-1}=\frac{5.442}{ln\left[3N(0)/N(1)\right]},
\end{equation} 

\noindent where $h\nu$ is the energy separation between rotational levels 
$N^{\prime\prime}$=0 and $N^{\prime\prime}$=1, while $g$ and $N$ are the 
statistical weight and column density
for each level. The derived CN excitation temperatures for these eleven 
components range from 2.55 K to 2.89 K with an average of 2.75 $\pm$ 0.10 K. 
The derived excitation temperatures indicate that no significant excitation 
in addition to that due to the CMB is observed in these cloudlets, including 
three components toward HD~204827 where very strong CN absorption is 
detected and $T_{ex}$ is between 2.72 and 2.78 K.

 The $b$-values in Table 4 for CO toward Cep OB2 and OB3 are most like 
those for CN, suggesting that these species coexist in the diffuse
molecular gas probed by optical absorption.  There is also a 
hint that the $b$-values for Cep OB3 are larger, as seen in other 
species, but in the case of CO, the difference could be the result of 
coarser resolution in $FUSE$ spectra for the stars in Cep OB3.  Other 
evidence makes the connection between CO and CN stronger.  First, 
while the input for the synthesis of CO bands was based on the 
component structure seen in CH, the best fit revealed a component 
structure very similar to that for CN. This is illustrated in Table 5, 
where the $rms$ deviations between CO and CH relative column density
fractions on the one hand 
and between CO and CN on the other are presented for syntheses based on 
STIS spectra. Along each line of sight, the $rms$ deviations are smaller 
for CN than for CH.  The correspondence is not perfect, however, 
because CO and CH are detected in cases where only upper limits for 
CN are available (e.g., Crenny \& Federman 2004).  Second, there is 
an approximate linear relationship between $N$(CO) and $N$(CN) for 
individual velocity components.  Since upper limits exist for both 
quantities, correlations were sought through the use of the package 
ASURV, revision 1.1 (see Isobe, Feigelson, \& Nelson 1986; Isobe \& 
Feigelson 1990; La Valley, Isobe, \& Feigelson 1992).  Using Schmitt's 
Method, we obtain log[$N$(CO)] $=$ ($1.16 \pm 0.13$) log[$N$(CN)] $+$ 
($0.75 \pm 1.59$).  This analysis suggests that all but 3 of 24 limits 
are consistent with detections.  When all data are treated as 
detections, Schmitt's Method gives a slope of $1.35 \pm 0.17$ and an 
intercept of $-$($1.56 \pm 2.02$) and a linear least-squares fit 
indicates respective values of $1.63 \pm 0.16$ and 
$-$($4.89 \pm 1.85$).  The slopes are steeper when all data are 
considered detections because the upper limits on CN are less 
confining.  Each of these observational effects show that columns 
of CO and CN are tightly coupled in the denser portions of diffuse 
molecular clouds, a conclusion consistent with our earlier 
findings (Zsarg\'{o} \& Federman 2003; Crenny \& Federman 2004).
   
\section {Correlations among Column Densities}

Comparisons of column densities for different species provide constraints on 
chemical models, elemental depletions, species distributions, and the 
processes contributing to the ionization in diffuse clouds (e.g., 
Hobbs 1976; Chaffee \& White 1982; Federman et al. 1994; Welty \& Hobbs 2001).
Based on high resolution and high S/N spectra, Paper I derived column 
densities for CN, CH, CH$^{+}$, \ion{Ca}{1}, \ion{K}{1}, and \ion{Ca}{2} for 
individual velocity components along 29 lines of sight. 
The data set has good internal 
self-consistency and is also large enough to allow us to discuss correlations 
based on individual components rather than 
total line of sight columns as usually done in previous studies (e.g., Danks, 
Federman, \& Lambert 1984; Welty et al. 2003). Therefore, presumably, this 
data set will provide a good view of relationships among species. 
We examine some possible correlations 
between column densities of observed species. In each case, we first 
perform a least-squares fit to log[$N$(Y)] versus log[$N$(X)] to obtain 
the linear correlation coefficient, $r$, and the corresponding probability 
of no correlation, $p$, --- i.e., the probability for a random distribution of 
N measurements to result in a correlation coefficient $r^{\prime} \ge r$, 
according to Bevington \& Robinson (1992). Then, a regression 
analysis\footnote{We used the subroutine regrwt.f, obtained from the Penn
State statistical software archive at http://www.astro.psu.edu/statcodes,
with slight modifications, to perform the regressions.} is
performed to get the slope, intercept, and an estimate of the scatter (the
root mean square distance of points from the fit line). In all the fits,
we made one 3$\sigma$ pass through the data to eliminate outliers. It turns
out that three points are excluded from the fit for the relationship of 
log[$N$(CN)] versus log[$N$(CH$^+$-corrected CH)] and one from fits of
CN vs. CH, \ion{Ca}{1} vs. \ion{K}{1}, and \ion{Ca}{1} vs. \ion{Ca}{2}. 
Table 6 provides a summary of the correlations. 
The analysis shows that all correlations listed in the table have confidence
levels, 1~$-$~$p$, exceeding 99.95\%. 
In the following subsections, we describe some of these relationships. 

\subsection{CH versus CH$^+$}

When all CH velocity components are considered, the data set
suggests a weak, at most, trend of increasing CH abundance with increasing 
CH$^+$ column. However, the correlation becomes much stronger when we include 
only CH$^+$-like CH components, as shown in the bottom panel of Figure 1.
A closer look reveals that the correction
is weak for components with higher CH$^+$ columns. If only components with
log[$N$(CH$^+$)] $>$ 12.6 are considered, column densities of the
two species are not well correlated. This is clearly indicated by the
greater amount of dispersion seen at higher column densities for a linear
fit to all the data. Therefore, the correlation is sought for
components with log[$N$(CH$^+$)] $\le$ 12.6. 

In previous studies, based on total column densities along lines
of sight, CH column densities were sometimes found well 
correlated with $N$(CH$^+$).
For instance, Crane, Lambert, \& Sheffer (1995) found that lines of sight 
with detectable amounts of both CH and 
CH$^+$ show a large range in CH to CH$^+$ column density ratios, whereas 
Federman, Welty, \& Cardelli (1998) showed that, for lines of sight with no 
CN detections,  CH was well correlated with CH$^+$, and that the CH 
predicted by CH$^+$ synthesis 
agreed well with observations. Our findings explain why different
correlations were seen in these studies. Because many 
of the sight lines in the study of Crane et al. have large amounts of 
CN, some CH on these sight lines is associated with CN instead 
of CH$^+$. Therefore, $N$(CH) should not be well correlated with
$N$(CH$^+$). On the other hand, Federman et al. (1998) focused on lines
of sight with detected CH and CH$^+$ but no corresponding CN.
All CH in this study was CH$^+$-like CH and a good correlation is expected. 

\subsection{CN Versus CH}   

The top panel of Figure 1 reveals that column densities of CN components are 
correlated with columns of their corresponding CH components (one 
component of HD~217312 was eliminated by a 3$\sigma$ pass). This 
component-based correlation is much tighter than the sight-line-based 
relationship (cf. Figure 7 of Federman et al. 1994). Moreover, the 
correlation is even stronger when only CN-like CH components are included, 
as shown in the bottom panel of Figure 2. This study strengthens our
previous statement, based on $b$-values, that 
CN-like CH components are closely associated with the CN components. 

We have discussed CN-like CH and CH$^+$-like CH.  
How about CH components that have both corresponding 
CN and CH$^+$ components? Is a part of their CH associated with CN, and 
another part with CH$^+$? To seek answers to these questions, we calculated 
the amount of CH corresponding to CH$^+$ for these components by applying 
the $N$(CH$^+$-like CH) vs. $N$(CH) relation given in Table 6, and 
subtracted the result from the total CH column density, yeilding 
$N$(CH$^+$-corrected CH). [If log[$N$(CH$^+$)] $>$ 12.6, 
$N$(CH) = 4.0 $\times~10^{12}$ cm$^{-2}$ is taken 
from the total CH column as CH$^+$-like CH]. 
The top panel of Figure 2 presents the correlation between 
$N$(CN) and $N$(CH$^+$-corrected~CH), where three components are 
excluded from the final fitting by the 3$\sigma$ pass. All three 
components have small amounts of CN, 
(0.4\sbond 0.7) $\times~10^{12}$ cm$^{-2}$, but
are still too much for their corresponding CH$^+$-corrected CH in
the relationship. It is possible that
their amounts of CN are overestimated because columns of about 
0.4 $\times~10^{12}$ cm $^{-2}$ represent marginal detections. As one can 
see from Figure 1, the 
$N$(CH$^+$-like CH) vs. $N$(CH$^+$) relationship shows quite a bit of 
scatter. This scatter may cause  the three components 
to be outliers as well. 
 
Use of CH$^+$-corrected CH interestingly does improve the correlation 
compared to the
original $N$(CN) vs. $N$(CH) plot. The correlation coefficient slightly 
increases from 0.80 to 0.86. As one can see from Figure 2 and 
Table 6, the  regression is now very similar to that for $N$(CN) vs. 
$N$(CN-like CH). This analysis indicates that our assumption is a reasonable 
one. In other words, we conclude 
that there are two kinds of CH in diffuse molecular clouds: CN-like CH  
and CH$^+$-like CH. 
Disentangling the amount of CH in each appears possible when utilizing
the relation between columns of CH$^+$-like CH and CH$^+$. In the past, only 
ultra-high resolution observations toward $\zeta$ Oph (Lambert et al. 1990; 
Crawford 1997) distinguished between these two possibilities.

\subsection {\ion{K}{1} Versus CH}

Welty \& Hobbs (2001) found an essentially linear relationship between 
$N$(\ion{K}{1}) and $N$(CH) by using total column densities along lines 
of sight. Based on column densities for individual velocity components, we 
obtain a very similar correlation, with a slope of 0.96 $\pm$ 0.04, as 
shown in the upper panel of Figure 3.  
Three facts suggest that these two species respond very similarly to changes 
in physical conditions in diffuse molecular clouds: 1) the strong linear 
correlation between $N$(K I) and $N$(CH), 2) the similar 
average $b$-values (Table 4) 
and 3) the similar column density relationship for components and for lines of 
sight.  This also implies that they generally coexist, at least in the density 
range sampled by these data. However, we also note
that there are exceptions. A few \ion{K}{1} components with reasonable 
strengths do not have corresponding CH components. For instance, 
the velocity component at $V_{LSR} \sim -$ 4.3 km s$^{-1}$ toward
HD 216532 has a \ion{K}{1} column density of 3.2 $\times$ 10$^{11}$ cm$^{-2}$,
but no detected CH. However, there are relatively strong \ion{Ca}{1} and
\ion{Ca}{2} lines with column densities of  5.2 $\times$ 10$^{9}$ cm$^{-2}$ and
10.0 $\times$ 10$^{11}$ cm$^{-2}$, respectively. Considering the $b$-value
of the \ion{K}{1} component, 1.5 km s$^{-1}$, is greater than the average 
value of 0.92 km s$^{-1}$ (as is the case for the \ion{Ca}{1} and \ion{Ca}{2} 
component),  we think
this \ion{K}{1} component may arise from a slightly lower density region where
no significant amount of CH is present, but one with
a relatively large volume so that a reasonably strong \ion{K}{1} line is
observed. In other words, this may indicate 
that \ion{K}{1} could be distributed in less dense regions where no observable
amount of CH resides.  

\subsection{\ion{Ca}{1} Versus \ion{K}{1} }

The lower panel of Figure 3 presents the relationship between 
$N$(\ion{Ca}{1}) and $N$(\ion{K}{1}). The figure shows that the column 
densities of the two 
species are well correlated. However, the best fit slope of 0.60 $\pm$ 0.04, 
indicated by the solid line in the plot, is significantly smaller than 1.0. 
Compared with the relationship
of Welty et al. (2003), which was based on total column densities,
our best fit slope is the same as theirs (0.60 $\pm$ 0.08) while the scatter 
and the intercept (2.88 $\pm$ 0.47 versus 3.11 $\pm$ 0.68) are slightly smaller. 
Welty et al. (2003) discussed possibilities that could lead to a slope 
less than unity for the comparison between log[$N$(\ion{Ca}{1})] and 
log[$N$(\ion{K}{1})]. They concluded that it is most likely 
that calcium depletion has a steeper dependence on local density than does 
potassium depletion. This is not all that unexpected because calcium is a
refractory element while potassium is more volatile, having lower
condensation temperature. Once again, three 
facts, the good correspondence in column density,  
similar average $b$-values (see Table 4), and the similar correlations 
between relationships based on individual components 
and on lines of sight, suggest that these two species generally share  
comparable volumes.   

In addition to above the correlations, Table 6 also lists relationships for
log[$N$(\ion{Ca}{1})] vs. log[$N$(\ion{Ca}{2})] and log[$N$(\ion{K}{1})] 
vs. log[$N$(\ion{Ca}{2})]. Column densities of these species are 
correlated although not as tightly as the relationships above,
with about twice the scatter. The column density ratios can
differ by a factor of 10 for individual components. Such a large range in
column density ratios may indicate that these species do not track each other 
well. In other words, they seem to be probing different portions of a cloud. 
There are actually many \ion{Ca}{2} components without corresponding 
\ion{Ca}{1} or \ion{K}{1}. This latter fact arises because \ion{Ca}{2} is 
believed to be distributed more broadly (Welty et al. 2003).

\section{Chemical Analysis}

\subsection{CN Chemistry}

The route to CN in diffuse molecular clouds involves CH to C$_2$ to CN 
(Federman et al. 1984, 1994).  We use the steady-state analytical expressions 
in Federman et al. (1994), with updated rate coefficients (Knauth et al. 
2001; Pan, Federman, \& Welty 2001), to extract estimates for gas 
density in the material containing CN by matching the observed column
densities for CN and C$_2$ (when available). Steady state is appropriate
for this scheme because the photochemical time scales are less than
100 yrs., much shorter than the sound-crossing time. 
$Observed$ CH column densities are adopted for use in estimating 
the CN and C$_2$ column densities.  In our earlier papers (i.e., Federman et 
al. 1984, 1994) the analysis is based on line-of-sight column 
densities, but here we obtain results for individual velocity 
components where CN is detected.  The CH and CN column densities 
mainly come from the results presented in Paper I.  The 
exception is CH toward $\lambda$ Cep; we adopt the results from 
the ultra-high resolution measurements of Crane et al.  
(1995).  (In passing, we note that the 2 CH components have a 
separation consistent with what we found for CN.  Of the 2 CH$^+$ 
components, only the bluer one is associated with gas containing CN; 
the redder component is seen in \ion{K}{1} absorption but not in 
absorption from other molecules $-$ see Paper I.)  Along several 
sight lines, results for C$_2$ are also available (Danks \& Lambert 
1983; Federman \& Lambert 1988; Federman et al. 1994).  When more 
than one CN component is present for a given direction, the relative 
amounts of C$_2$ in each component is taken to be the same as CN 
because these molecules likely coexist (Federman et al. 1994).

To extract estimates of gas densities from the chemical model (Federman
et al. 1994, Pan et al. 2001), in addition to column densities of
CN and CH (C$_2$ column provides a further constraint), 
O, C$^+$, and N abundances, temperature, and UV radiation field are
needed. Our sample contains some of the most molecular-rich diffuse clouds 
studied to date.  Special care is necessary when choosing the 
appropriate atomic abundances in these circumstances.  The O and 
C$^+$ abundances are now available along sight lines in our 
sample. Cartledge et al. (2001) found that the O abundance toward 
$\rho$~Oph~D and HD~207198 differs by about 1-$\sigma$ from the 
diffuse cloud average, while that for C$^+$ is comparable to the 
average (Sofia et al. 2004). Andr\'{e} et al. (2003) and Cartledge et al. 
(2004), respectively, provided results on O toward HD~206773 and
$\lambda$ Cep that are not atypical. Thus, we continue to use the average 
abundance for diffuse clouds in the chemical analysis.  For C$^+$ 
this is not unreasonable because the largest CO columns 
encountered in our survey are about 10$^{16}$ cm$^{-2}$, which 
represents less than 10\% of the carbon budget, and the neutral 
carbon column densities (Jenkins \& Tripp 2001) are smaller still. 
Most calculations are based on the average interstellar radiation
field of Draine (1978) and on $T$ $=$ 50 K,
the temperature commonly inferred from excitation of low-lying
rotational levels in H$_2$ and C$_2$ (e.g., van Dishoeck \& Black
1986; Federman et al. 1994). The latter value is not critical because the 
results for $n$ are not very sensitive to $T$ --- the estimated gas densities
change by less than 10\% when the temperature is changed by
a factor of 2 (Pan et al. 2001).  
For especially molecular-rich clouds, 
lower values for $T$ are adopted, while slightly larger temperatures (65 K) 
are used for sight lines toward Cep OB3 where fewer CN components are 
present. These adopted temperatures and the UV radiation field are
justified in $\S$6.2. 

The chemical model also involves the optical depths of grains at 1000 \AA,~
$\tau_{uv}$, and threshold grain optical depths for ultraviolet photons, 
$\tau_{uv}^{\rm o}$, in somewhat indirect ways. These optical depths were
used to determine the amount that photodissociation rates are attenuated
and a coarse treatment of the C$^+$ to CO transition, respectively. For 
most sight lines, we previously 
adopted $\tau_{uv}$ $=$ 2$A_V$ (e.g., Federman 
et al. 1994), but such a value does not appear to be appropriate 
for our sample of sight lines.  As noted by Federman et al. (1994), 
larger grains are present toward the stars comprising $\rho$ Oph, 
and a value of 1.4$A_V$ is used.  Though the extinction law for 
directions in Cep OB3 is similar to the law representing the average 
for diffuse sight lines, the molecular gas is more clumpy (Sargent 1979) 
and Federman et al. (1994) adopted 1.4$A_V$ here as well. We do the 
same in light of the numerous components seen in \ion{K}{1} 
absorption.  For Cep OB2, extinction laws for several stars are 
available (Massa, Savage, \& Fitzpatrick 1983; Savage et al. 1985; 
Aiello et al. 1988; Fitzpartick \& Massa 1990; Megier et al. 1997), 
although sometimes only in tabular form.  For the most part, the 
laws do not differ much from the average law, except for the sight 
line very rich in molecules $-$ HD~204827, where the ultraviolet 
extinction rises quite rapidly at the shortest wavelengths (Fitzpatrick 
\& Massa 1990).  Our starting point for the directions in 
Cep OB2 for $\tau_{uv}$ was 2$A_V$.  The number of components toward 
stars in this association as well as in Cep OB3 created a further 
complication, which is compounded by the fact that foreground 
material along the $\approx$ 750-pc pathlengths is likely.  First, 
we assumed that all the molecular gas is located in the star-forming 
regions.  Extinction only from the molecular components was 
considered in the chemical analysis; it was estimated from the 
fraction of \ion{K}{1} column associated with CH 
components.  The fraction always exceeded 0.50 and was usually 0.80 
or more. Second, the line-of-sight extinction was used because 
shadowing of one diffuse molecular cloud on another is more than 
likely.  Since photodissociation is the dominant destruction pathway 
for most of the clouds in our study, uncertainties in $\tau_{uv}$ 
lead to uncertainties of about 30\% in the inferred gas densities.
Since the directions under study have a significant amount of 
reddening [$E$($B-V$) $\ge$ 0.3 mag], but diffuse cloud abundances 
for C$^+$, the perscription of Federman \& Huntress (1989) 
for treating the C$^+$ to CO transition had to be revisited.  The 
cause of the apparent inconsistency between reddening and C$^+$ 
abundance is the presence of numerous molecular components along 
each direction in Cep OB2 and OB3. For these associations, the 
threshold grain optical depth for ultraviolet photons, 
$\tau_{uv}^{\rm o}$, was increased to 3.75 from 2.0.  As a result, 
only toward HD~204827, where $N$(CN) reaches the highest values, is 
there likely to be much conversion of C$^+$ into CO. 

The results of our analysis, which are based on the rate equations from
Federman et al. (1994) with updated rate data from Knauth et al. (2001)
and Pan et al. (2001), appear in Table 7. The simplified model cloud
has constant density and temperature. For each cloud along
a specific direction, we list the factor giving the enhancement 
over Draine's (1978) interstellar UV radiation field ($I_{uv}$), 
$\tau_{uv}$, the kinetic temperature ($T$), the gas density 
[$n$ $=$ $n$(H) $+$ 2$n$(H$_2$)], the observed column densities for 
CH, C$_2$, and CN [$N_o$(CH), $N_o$(C$_2$), and $N_o$(CN)], and 
the predicted columns for C$_2$ and CN that best match the 
observations [$N_p$(C$_2$) and $N_p$(CN)]. The column densities 
are given in units of 10$^{12}$ cm$^{-2}$. 
The predicted column densities are always within 30\% 
of the observed values; thereby providing another measure of the uncertainty
in inferred gas density.

The inferred gas densities range from about 100 to slightly higher 
than 1000 cm$^{-3}$, not untypical for diffuse molecular gas, and
consistent with those by Weikard et al. (1996) --- based on CO
observations, they estimated densities for CO cloudlets of about
500 to 2000 cm$^{-3}$.  Our inferred
densities probably represent lower limits to the true densities.  
First, we assumed that all the CH in a component participated in 
the production of CN although we find CN-like and CH$^+$-like CH 
(see Section 5.2).  Use of the relationship between CH$^+$-like CH 
and CH$^+$ to account for CH in the lower density gas along the 
line of sight and not involved in CN chemistry is possible.  We chose 
not to pursue this because the dispersion in the relationship 
(Fig. 1) could lead to unphysical results (negative columns) in 
some of our sample.  Instead, we note that the 
clouds with the lowest densities are the ones most likely affected 
by this assumption.  Second, we obtained the extinction for the 
molecular gas along the line of sight from the fraction of 
\ion{K}{1} in CH components.  The extinction would be smaller if 
only the CN components were used, resulting in higher densities to 
offset more photodissociation.

We analyzed the CN chemistry for several of these sight lines in 
the past (Federman \& Lambert 1988; Federman et al. 1994; Pan  
et al. 2001).  For the most part, the inferred gas 
densities are similar, usually within 30\%.  Much of the difference 
is likely caused by our current emphasis on individual components 
along a line of sight.  The other factor, highlighted by the 
comparison with the results of Pan et al. (2001) based on the same 
chemical rates, is the use of updated column densities.  For 
example, the difference in density for the $-0.8$ km s$^{-1}$ 
component toward HD~206267C arises from the decrease in $N$(CN) 
from $0.9 \times 10^{12}$ to $0.4 \times 10^{12}$ cm$^{-2}$.  

Two sight lines, HD~207198 and HD~204827, yield surprisingly low gas 
densities for the large CN columns present.  
For HD~207198, the three CN components contain 
substantial amounts of CH$^+$.  If the CH associated with CH$^+$ 
does not take part in CN production and therefore must be removed 
from the CH column used in the chemical model, then densities a 
factor of 2 larger would be inferred.  This modification is less 
effective for the clouds toward HD~204827.  Instead, the low values 
inferred for $n$ result from the significant extinction.  The large 
extinction, consistent with HD~204827 lying deepest within the 
CO contours of any sight line in our sample 
(see Fig. 8), greatly lessens the 
importance of destruction through photodissociation.  This contrasts 
with the results for HD~62542 where high densities are required to 
reproduce the large CN column on a sight line with $A_V$ of 1 mag 
(Cardelli et al. 1990; Federman et al. 1994).  These points are 
illustrated in Table 7 for the $+1.9$ km s$^{-1}$ component 
toward HD~207538.  The second entry shows that $n$ increases five-fold 
for $I_{uv}$ $=$ 5, highlighting the importance of photochemistry 
for most of our sample.  The third entry indicates that when about 
half the CH is associated with CN, $n$ increases a factor of 
$\sim$ 2.

\subsection { H$_2$ Excitation}

Column densities for H$_2$ rotational levels in the ground vibrational level 
carry information on physical conditions that provides a justification
for our inputs in the chemical analysis. The relative populations of 
$J$ $=$ 0 and 1 give a reliable estimate for kinetic temperature 
(Savage et al. 1977).  Since rotational transitions between 
ortho-H$_2$ (odd $J$) and para-H$_2$ (even $J$) are strictly 
forbidden, collisions involving protons and H$_2$ produce a thermal 
distribution between $J$ $=$ 0 and 1. High-J levels may 
be populated by photon pumping, collisions in shock-heated gas, and 
in the formation of H$_2$ molecules on dust grains.  Column density 
ratios such as $N$(4) and $N$(2) yield upper limits to the UV flux 
causing the photon pumping when the other processes are 
neglected (but see Gry et al. 2005).

In Table 2, we listed kinetic temperatures from the H$_2$ analysis ($\S$ 3)
for lines of sight toward Cep OB2 and Cep OB3. The listed values of 
$T_{1,0}$ suggest that directions toward Cep OB3 have slightly higher 
kinetic temperatures (80 K on average) than those toward Cep OB2 (74 K). 
The temperatures used in the chemical analysis ($\S$ 6.1) are somewhat 
lower, 65 K for Cep OB3 and 50 K for others, as would be expected for 
the less extensive, denser portions 
of a cloud where CN is detected.  As for higher J levels, the observed 
ratios, log[$N$(4)/$N$(2)], presented in Table 3 lie between $-$2.68 
and $-$0.95.  Placing our results for H$_2$ on Fig. 3 of Browning, 
Tumlinson, \& Shull (2003), which is 
based on $I_{uv}$ $\approx$ 15, allows us 
to infer the relevant UV flux to adopt. The data for HD~203374A and 
HD~208440 are to the left of the points in the figure in Browning et al., 
indicating fluxes consistent with the average interstellar value. On the 
other hand, the data for 9 Cep, HD~209339, and HD~217035A lie to 
the right, suggesting fluxes along these lines of sight are
higher than the average interstellar value.
An UV enhancement factor as large as 100 may be needed 
to explain our H$_2$ results. Because other processes leading to 
high-J level populations and because the presence of less shielded foreground 
gas that is mainly atomic were not considered, $I_{uv}$ values inferred
from H$_2$ data should be regarded as upper limits for the chemical
analysis ($\S$6.1). Furthermore, such large fluxes are usually associated
with correspondingly high densities and temperatures in photon dominated
regions (e.g., Knauth et al. 2001), but our CO data indicate subthermal
excitation in diffuse molecular gas. Therefore, our use of the average 
interstellar radiation field ($I_{uv}$ $\sim$ 1) 
seems reasonable under these circumstances along lines of sight with CN 
absorption. In other words, it appears that the material studied here is
sufficiently far from the background star ($\sim$ 3 pc) that only the
average interstellar field is important. 

For the 14 directions with H$_2$ measurements, knowledge of 
both the $b$-value for each component and the mean kinetic 
temperature along the line of sight yields an estimate for the 
turbulent velocity in each component. The turbulent velocity 
for all but one of the 72 components is $\approx$ 1 km 
s$^{-1}$.  The results for gas toward Cep OB2 and OB3 are 
indistinguishable.  Since the material along these sight 
lines is predominantly atomic, a comparison based on sound 
speeds derived from a ratio of specific heats of 5/3 is 
appropriate.  For temperatures of 70 to 80 K, the sound 
speed is also about 1 km s$^{-1}$.  Thus, the turbulence 
appears to be sonic.  One component toward $\lambda$ Cep is 
the exception: its turbulent velocity is $\sim$ 2.5 km 
s$^{-1}$.  The component resembles the CH$^+$-like CH 
component toward $\zeta$ Oph (Lambert et al. 1990; Crawford 
1997).

\section {Discussion}

\subsection{Distribution of Species}

In previous sections, similarities in average $b$-values and 
correlations between column densities allowed us to suggest that
some species coexist, while others probe different portions of a
diffuse molecular cloud with typical visual extinctions of 1 to 2
mag. We here investigate the distribution of observed species from
other aspects.  
If two species coexist, their spectral profiles along a given line of sight 
should appear similar, although their line strengths may be quite different. 
For easy comparison, we construct apparent optical depth (AOD) 
profiles (Savage \& Sembach 1991) for the 
absorption lines, and overplot scaled AOD profiles for different species on 
a given line of sight.
Figure 4 shows the comparison of these profiles along the line of sight toward 
HD~207308. The scaling factors differ from species to
species and from panel to panel. For instance, CH and \ion{K}{1} AOD
profiles are scaled by factors of 1/2.84 and 1/15.1 in the middle panel, 
respectively. 

By comparing AOD profiles for six species on each of our sight 
lines, we find, along a given sight line, (1) that the CN profile is the 
narrowest one, whereas the \ion{Ca}{2}
profile is the widest one, (2) that CH and \ion{K}{1} profiles are usually 
similar in both width and shape, and (3) that in most cases, CH$^+$ and 
\ion{Ca}{1} profiles are wider than CH, in roughly 80\% of the cases for 
CH$^+$, and 60\% for \ion{Ca}{1}. The similarity among 
the profiles for the three species is not all that clear. In some
cases, \ion{Ca}{1} profiles are similar to CH$^+$ ones. In other cases, 
they may be more similar to those of CH than to those of CH$^+$. It may 
depend upon the local gas density. Here we are assuming line width,
or equivalently $b$-value, is a measure of the distribution along the
line of sight. 

The comparison of AOD profiles suggests that \ion{Ca}{2} is the 
most widely distributed among the six species, followed by CH$^+$ and 
\ion{Ca}{1}, then \ion{K}{1} and CH, and finally CN. Distributions of 
velocity components reinforces this statement. Figure
5 is a histogram plot for distributions of velocity components with 
respect to $V_{LSR}$ along lines of sight in Cep OB2 (excluding components 
toward HD~206267A and $\lambda$ Cep because component structures are not
available for some species). The Figure shows 
that \ion{Ca}{2} components are much more widely distributed than those of 
other species, and that distributions 
of CH, \ion{K}{1}, and \ion{Ca}{1} have very similar shape. This  
again indicates correlations among the latter three species.

Previous studies of diffuse molecular clouds (Federman et al. 1984) showed 
that the CN molecule is produced in observable quantities only after 
significant amounts of precursor molecules, such as CH and C$_2$, are 
available. In particular, the relationship between $N$(CN) and $N$(H$_2$)
has a slope much steeper than that seen for $N$(CH) vs. $N$(H$_2$) (e.g.,
Federman 1982; Danks et al. 1984). Federman et al. (1984) attributed the
steepness in slope to the number of chemical steps before the molecule can be
detected. Higher densities facilitate the transformation along
the chemical sequence.  
Our observations reveal that CN has the narrowest profile, 
smallest $b$-values, and smallest number of velocity components among the 
species. In other words, CN is the least widely 
distributed and occupies the smallest volume among the species. Therefore, 
we believe that CN mainly resides only in denser regions of diffuse 
molecular clouds,  and that no observable amount of CN is present in low 
density clouds or in cloud envelopes. The correspondence between CO and CN
noted above suggests, for the most part, that the same applies to CO. This
is consistent with the slope found for $N$(CO) vs. $N$(H$_2$) --- 
see $\S$ 7.4; CO can exist in regions with somewhat lower densities because 
faster (ion-molecule) reactions dominate its production (e.g., 
van Dishoeck \& Black 1986). 

In $\S\S$ 4 and 5, we showed that there are two kinds of CH-containing material: 
CN-like CH, which is associated with CN production, and CH$^+$-like CH, 
which is related to the formation of CH$^+$. This suggests that CH 
could coexist with either CN and CH$^+$. Since in all velocity components 
where CN is detected,  CH is observed, we believe that CH exists in the 
whole volume where CN resides. On the other
hand, generally wider profiles, more components, and larger $b$-values
indicate that CH occupies a larger volume than does CN. 

Similarities in $b$-values, AOD profiles, and a good 
correlation between their column densities all suggest that 
CH and \ion{K}{1} generally coexist, or share a large fraction of 
their volumes in diffuse clouds.
The combined set of facts leads us to
believe that these two species can also exist in less dense gas, or less 
dense regions of clouds (such as envelopes) where CN is not present. A 
relevant question 
is to what extent do the CH and \ion{K}{1} distributions extend into the 
region of lower
density. In other words, what is the lowest gas density that can be
probed by CH and \ion{K}{1} observations? The
relatively low ionization potential (4.341 eV) for \ion{K}{1} suggests that 
it cannot exist 
in a very low density environment because it must be shielded from
photoionization. In addition, the $b$-values of CH and \ion{K}{1} are only 
slightly larger than that of CN. This indicates that their distributions 
cannot extend much
beyond the region containing CN. Therefore, CH and \ion{K}{1} appear to be
distributed in high- and moderately high-density regions of diffuse clouds 
($n \ge$ 30 cm$^{-3}$). We also note that K may be depleted 
onto grains in very high density gas, and that there are some \ion{K}{1} 
components without corresponding
CH components. The facts suggest that \ion{K}{1} may be absent in extremely 
dense regions of clouds, but slightly extend to less dense regions compared
to CH (also see $\S$ 5.3).

Because the AOD profiles for CH$^+$ and \ion{Ca}{1} are generally 
wider than those for CH and \ion{K}{1}, these species may occupy bigger 
volumes of a cloud. On the other hand, similar $b$-values and strong
correlations between $N$(\ion{K}{1}) and $N$(\ion{Ca}{1}) suggest that
\ion{K}{1} and \ion{Ca}{1} share a large fraction of their volumes. 
According to the chemical models of
Federman (1982) and Danks et al. (1984) and the depletion characteristics
of calcium (see next section), 
both CH$^+$ and \ion{Ca}{1} cannot have a 
significant abundance in dense gas where CN and CH reside. 
CH$^+$ and \ion{Ca}{1} only exist in less dense gas, and so
their distributions extend throughout the envelope of a cloud beyond those 
of CH and \ion{K}{1}. This is also consistent with the fact that \ion{Ca}{1} 
has a greater first ionization
potential than does \ion{K}{1} (6.1 eV vs. 4.3 eV). On the other hand, neither 
CH$^+$ nor \ion{Ca}{1} can exist in a very low density enviroment where 
insignificant amounts of H$_2$ (e.g., Federman 1982) are available for 
CH$^+$ production and where the main forms of Ca will be \ion{Ca}{2} or 
\ion{Ca}{3}. Thus, we conclude that CH$^+$ and \ion{Ca}{1} mainly 
exist in moderately high- to intermediate-density gas 
($n \sim$ 10--300 cm$^{-3}$).

The \ion{Ca}{2} ion is the most widely distributed species among the species 
we observed. Its absorption can arise in components not seen in neutral
atoms or molecules (see Figs. 4 and 5), and its $b$-value is the largest among
our observed species. Considering that Ca readily 
depletes onto grains, \ion{Ca}{2} should exist in a moderately high 
to low-density enviroment. Figure 6 is a schematic showing the distributions 
of species we discussed. In some low density clouds, the density is so low 
that molecules are not present even in the densest regions. \ion{Ca}{2} may 
be the only species among those studied here in such clouds. 
  
\subsection{Ca and K Depletions}

Cardelli, Federman \& Smith (1991) noticed that the \ion{Ca}{1} 
absorption line profiles
were more similar to those of CH$^+$ than to those of CH on some lines of 
sight and suggested that calcium depletion depends on the density.
For total column density along a line of sight, they found an inverse 
linear relationship between the ratios $N$(CN)/$N$(CH) and 
$N$(\ion{Ca}{1})/$N$(CH). Using a simple model,
assuming photoionization equilibrium in predominantly neutral gas, where 
carbon is the primary source of electrons and where the space 
densities and column densities have the same functional form, they concluded 
that the fractional density of calcium varies roughly as $n^{-3}$,
due to the depletion. However, Welty et al. (2003) recently found a
steeper relationship between the column density ratios with a slope of 
$\lesssim -$2.5, implying that the 
density dependence of Ca column density is not steeper than $n^{-1.8}$.  

Comparing our spectra, we found some cases where the \ion{Ca}{1} profiles 
are similar to those of CH$^+$ and in other cases the similarity is with
CH profiles. A closer examination shows that the similarities may
depend on the local gas density where the absorption arises. In most 
(roughly 75\%) of the high density cases [relatively large ratio of 
$W_{\lambda}$(CN)/$W_{\lambda}$(CH) or strong CN absorption], \ion{Ca}{1}  
profiles resemble those of CH$^+$. For intermediate density 
clouds [small ratio of $W_{\lambda}$(CN)/$W_{\lambda}$(CH) or no CN 
detection], we did not find a clear trend, but we note that in
some cases, the \ion{Ca}{1}, CH$^+$, and CH profiles are quite similar.
These findings on profiles are consistent with the picture for the distribution
of species drawn in $\S$~7.1. In a high $n$ enviroment, a large fraction 
of CH absorbers come from a region where no detectable amount of \ion{Ca}{1} 
and CH$^+$ reside, whereas along the line of sight a large fraction of 
\ion{Ca}{1} and CH$^+$ absorbers originate from a common volume. Then,
\ion{Ca}{1} profiles are more similar to those of CH$^+$ than those of CH. 
For a intermediate density cloud, the three species may coexist over a 
large fraction of the volume, and their profiles would be similar. Therefore,
qualitatively, the calcium depletion depends upon the local gas density.

As already noted, using total column densities, Cardelli et al. (1991) and 
Welty et al. (2003) obtained different slopes for the relationship between 
$N$(CN)/$N$(CH) and $N$(\ion{Ca}{1})/$N$(CH).  Figure 7 presents relations 
between the ratios $N$(CN)/$N$(CH) and both $N$(\ion{Ca}{1})/$N$(CH) 
and $N$(\ion{K}{1})/$N$(CH) for individual components. Fair correlations
are found between the ratios with correlation coefficients of $-$0.68 and
$-$0.40 and corresponding correlation confidences of 99.98\% and 99.50\%,
respectively.
Regression fits yield slopes of $-$1.03 $\pm$ 0.15 and 
$-$2.03 $\pm$ 0.20 for the respective relationships. Within the framework
of Cardelli et al., these slopes imply that the calcium and potassium 
column densities are proportional to $n^{-2.9}$ and $n^{-2.0}$, respectively,
due to their depletions. 
However, caution should be exercised in applying the framework because 
assumptions may not be valid. These species do not occupy exactly 
the same volume in a cloud, as stated in $\S$ 7.1, and  Welty et al. (2003)
argued that the photoionization equilibrium may not strictly hold in
diffuse clouds. While the inferred density 
dependence due to depletion may not be very accurate, it appears that 
calcium depletion varies more steeply on local density than potassium 
does, as suggested by both this analysis and by the slope in Figure 3. 

\subsection{Cloud Structure}
\subsubsection{Large-scale Structure}

The Cepheus bubble, at a distance of $\approx$ 750 pc from the Sun, has been the 
subject of many investigations. Radio \ion{H}{1} and CO observations revealed
that the bubble has structures on a scale of a few pc.
Expanding shells of dense gas around 
\ion{H}{2} regions have been 
inferred in previous studies of CO emission from the molecular clouds around 
OB stars. The large-scale CO molecular cloud shows a clumpy and irregular 
structure, with small cloudlets having sizes of a few parsecs (e.g., Patel et 
al. 1995, 1998; Weikard et al. 1996). Figure 8 is the map of CO emission for
the Cep OB2 region adapted from Patel et al. (1995; 1998), with our stars 
projected onto the map. 
According to their locations on the CO map, we divided our
stars into several groups (that do not reflect coeval stellar 
groups). For example, we consider HD~206267A, C, D, and HD 206773 as one 
group. They are associated with the same CO cloud, cloud number 18 in 
Table 2 of Patel et al. (1998). HD 207538, HD 208266, and $\nu$ Cep are in 
another group associated with cloud number 21. The 
AOD profiles for absorption lines are
very similar for a given species in a group. The lines of sight in each group
have most velocity components in common, especially within the $V_{LSR}$ 
range in which CO emission was detected from a cloudlet. Spectra for sight 
lines passing near edges of CO clouds show strong CN absorption.
However, there are some exceptions. For example, the line of sight 
toward 14 Cep passes near the edge of
one CO cloud, but only one weak CN component with $N$(CN) = 
0.45 $\times$ 10$^{12}$ cm$^{-2}$ was detected. There is no nearby
CO cloud toward HD~207198, but strong CN and CO
absorption was detected on the line of sight.  
Since 14 Cep has the smallest extinction 
[$E$($B-V$)=0.35] and the smallest estimated distance (650 pc) among our 
lines of sight in
Cep OB2, 14 Cep most likely is located in front of its nearby CO 
cloudlet. In general, our data are consistent 
with the large-scale structure suggested by maps of CO radio emission.
The $V_{LSR}$ of CN , CO, and CH components on a line of sight 
(including those toward $\rho$ Oph and Cep OB3) are in a range within which 
CO emission is detected from a cloudlet. For IC 1396 in Cep OB2, this 
correspondence is consistent with the conclusion of Weikard et al. (1996)
that much of the material seen in CO emission is in front of HD~206267.
The sensitivity of absorption measurements at visible and UV wavelengths
allows us to detect smaller molecular columns than is possible through
emission line maps, providing further detail of cloud dispersal in regions
of star formation. 

\subsubsection{Small-scale Structure}
While it is well established that the column densities of some atomic 
(\ion{H}{1}, \ion{Na}{1}, \ion{K}{1}) and  molecular (H$_{2}$CO, 
OH) species vary on scales of 10 to 10$^{5}$ AU along diffuse sight lines,
detailed information on the physical conditions causing the 
variations is generally not available. A key question posed by the 
observations is whether the subparsec structure is caused by density 
variations (e.g., Frail et al. 1994; Lauroesch \& Meyer 1999; Crawford 
et al. 2000), by fluctuations in ionization equilibrium 
(Lauroesch \& Meyer 1999; Welty \& Fitzpartrick 2001), 
by the geometric structure of clouds (Heiles 1997), or by something else. 
A crucial point is to determine accurate physical conditions for the gas 
showing variations in column density. With gas densities inferred from a 
chemical model for CH, C$_{2}$, and CN (see $\S$ 6.1) for velocity components 
on lines of sight toward members of multiple star systems, we now discuss 
small scale structure in diffuse molecular gas.

There are three multiple star systems, $\rho$ Oph~A/B/C/D, HD 206267~A/C/D, 
and HD 217035~A/B, in our data set. 
Separations between members of systems range from 450 to 38,320 AU (0.02 to
0.18 pc). Substantial variations in CN absorption are observed among sight 
lines of $\rho$ Oph. There are striking differences in the CN, CH, CH$^+$, 
\ion{Ca}{1}, and \ion{K}{1}
profiles among the three sight lines of HD 206267. No CN is detected toward 
either line of sight in the HD 217035 system, but a significant difference 
in the CH$^+$ profiles for the two members of the system is observed. The 
corresponding column densities differ by a factor of $\sim$2 in 
individual velocity components.  

Pan et al. (2001) discussed gas density variations on sight lines
of the multiple star systems HD~206267 and HD~217035.
Because Paper I employed a different method to determine column densities of
individual components and a slightly different data reduction procedure,
velocity component structures and their column densities might be slightly
different from those in Pan et al. However, the conclusion of Pan et al. that
the component gas densities can differ by factors of 5 between adjacent
lines of sight is not altered. From Table 7, we see that the derived gas 
densities of the components toward members of HD 206267 at the same or
similar $V_{LSR}$ differ by factors of 2\sbond 7. (If the upper 
limits for gas density are considered, the difference can be as high as
10.) In the case of the $\rho$~Oph multiple star system, 
the diffuse molecular cloud toward $\rho$~Oph~C has a density about 
twice that inferred along the other three nearby sight lines. While 
less dramatic than the results for the system HD~206267,
the result for $\rho$~Oph reinforces the conclusions of Pan et al.  
Density contrasts up to factors of 10 over scale lengths 
of 10,000 to 20,000 AU {\it are present} in the densest 
portions of diffuse clouds, those sampled by CN absorption; 
variations of the order of 10$^4$ suggested by 21 cm observations 
(e.g., Frail et al. 1994) must have a different origin.  
Possibilities include fluctuations in ionization equilibrium 
(Lauroesch \& Meyer 1999; Welty \& Fitzpatrick 2001) and the 
geometric structure of clouds (Heiles 1997).  The smallest scale 
length probed by our measurements, 450 AU between $\rho$~Oph~A and B, 
shows no variation in modeled density greater than about 35\%.

Cloud thicknesses for individual components (cloudlets) along
the line of sight toward the background star can be 
estimated as well. The total column density of protons, $N_{tot}$(H), 
for a component can be estimated from our measurements of $N$(\ion{K}{1})
(from Paper I) and the 
relationship between $N$(\ion{K}{1}) and $N_{tot}$(H), 
log[$N$(\ion{K}{1})] $= (-26.9 \pm 2.7) +(1.8 \pm 0.1)$ log[$N_{tot}$(H)]
derived by Welty \& Hobbs (2001), under the assumption that it applies to 
components although they used total column densities.  Then 
$n$ and $N_{tot}$(H) are used to estimate the thickness of a 
cloudlet.  For the multiple star systems, $\rho$ Oph and HD~206267, 
the thickness is within 50\% of 1 pc, except for the two low-density 
components toward HD~206267A where the thickness is about 5 pc.  
Since $n$ refers to the dense region where CN resides, the mean $n$ 
for the whole cloudlet should be lower.  Our estimates should then be 
regarded as lower limits.  If the lower limits are not far below 
their true values, they suggest that some of the cloudlets (in Cep 
OB2) are sheet-like with aspect ratios of 5 to 10 because 
maps of CO millimeter-wave emission show that a typical size of a 
cloudlet on the sky is a few parsecs (Patel et al. 1995, 1998; 
Weikard et al. 1996).  
Our results, therefore, provide evidence for the need to consider 
non-spherical geometries (Heiles 1997) 
when analyzing small scale structure in diffuse clouds.

\subsection{Individual Cloud Systems}

The diffuse gas associated with the 
three star-forming regions, Cep OB2, Cep OB3, and $\rho$ Oph
(belonging to Sco OB2), were observed. Table 8 lists the average number, 
{$\langle$M$\rangle$}, of 
velocity components per sight line for each species and the average column 
density of each component for each species.\footnote{CO column densities
for individual components were obtained along 15 lines of sight toward
Cep OB2 and Cep OB3 by analyzing {\it HST} and {\it FUSE} spectra --- see
$\S$ 3. Because different spectral resolutions comprise this sample, it may be 
not fully appropriate to include the CO results in the comparison. Still, our 
analyses show
that CO has fewer components than CH but more than CN,  and that, on average, 
sight lines toward Cep OB2 and Cep OB3 have about the same number of
CO components per line of sight, but components in Cep OB2 have much larger
mean column density.}  
The table shows that lines of sight toward Cep OB2 and
Cep OB3 have many more \ion{Ca}{2} and \ion{K}{1} components than those 
toward $\rho$ Oph. This is consistent with radio observations (Sargent 1977; 
Weikard et al. 1996; Patel et al. 1998), which revealed that clouds in 
Cep OB2 and Cep OB3 have complicated structures. Table 8 also
reveals that Cep OB3 has the greatest mean
number of \ion{Ca}{2} and CH$^+$ components per sight line, the largest 
average column density per \ion{Ca}{2} component, but the smallest 
corresponding values for CN among the three star-forming regions. This 
indicates that a larger fraction of low density material is probed
by the lines of sight toward Cep OB3. 
Because CN resides only in the denser region of diffuse molecular clouds, 
while \ion{Ca}{2} is widely distributed, the ratio, 
$\langle$M$_{{\rm CN}}\rangle \times\langle N$(CN)$\rangle$/
($\langle$M$_{{\rm CaII}}\rangle\times\langle N($\ion{Ca}{2})$\rangle$), 
may reflect the fraction of denser material in the clouds. The ratio for $\rho$
Oph is about 2 and 10 times larger than for Cep OB2 and Cep OB3, respectively. 

Lines of sight toward Cep OB3 have larger mean extinction, $E(B-V)$, 
(0.78 vs. 0.53 mag), greater $\langle$M$_{\rm {CaII}}\rangle$, and larger 
$\langle N$(\ion{Ca}{2})$\rangle$ 
compared to those toward Cep OB2. This suggests that more material is 
intercepted along lines of sight in Cep OB3. However, Cep OB3 clouds have 
a lower mean density. This indicates that Cep OB2 clouds are geometrically
thinner, along the line of sight, than Cep OB3 clouds. 

Using total H$_2$ column densities from {\it Copernicus} observations,
Federman (1982) and Federman et al. (1980) found that total CH and CO
column densities along lines of sight were well correlated with total
H$_2$ column densities. In the present study, we obtained total H$_2$
column densities along 15 sight lines toward Cep OB2 and Cep OB3 based
on $FUSE$ observations. Although $FUSE$ provides higher quality
spectra, we still could not extract reliable column 
densities for individual components due to very large line optical depths, 
thus preventing us from seeking correlations between $N$(CH) and $N$(CO) 
with $N$(H$_2$) on a component basis as we did for other relationships in
$\S$ 5. Figure 9 shows plots of total column density
for CH and CO versus H$_2$ along the 15 lines of sight.
In general, the Figure reveals the same trends, as previous studies (Federman
et al. 1980; Federman 1982; Danks et al. 1984): $N$(CH) 
and $N$(CO) increase
with increasing $N$(H$_2$) with a slope of 0.95 $\pm$ 0.10 for 
CH (Cep OB2) and 3.16 $\pm$ 0.34 (Cep OB2) and 2.93 $\pm$ 0.58 (Cep OB3)
for CO. For comparison, Danks et al. (1984) obtained a slope of
0.90 $\pm$ 0.10 for $N$(CH) vs. $N$(H$_2$), and Federman et al. (1980)
found a slope of about 2 for the CO trend. However, Cep OB2 sight lines 
are clearly distinguished from those toward Cep OB3.
Lines of sight toward Cep OB2 have higher $N$(CH) and $N$(CO)
column densities for a given $N$(H$_2$). This
implies that the physical conditions in these star-forming regions, such as
average gas density, may be quite different. Since there are a larger 
fraction of CN components toward stars in Cep OB2, the diffuse molecular 
clouds toward Cep OB2 have higher average gas density. 
The direction toward HD~217312 in Cep OB3 represents an intermediate case: 
there is a large enough fraction of CN components to place it among the CH
data for Cep OB2, but not enough for the CO data because CO traces the
densest gas with CN. 

  Absorption from the diffuse molecular gas for a specific star-formation
region shows similar overall structure, and the length scales probed by 
our measurements are similar for the three star-forming regions. However,
the diffuse molecular clouds associated with Cep OB3 exhibit 
different characteristics compared to those associated with $\rho$ Oph and 
Cep OB2. The cloudlets 
have lower gas density and lower total molecular mass (see $\S$ 2).
Compared with Cep OB2 and Cep OB3, material in the $\rho$ Oph star-forming
cloud is much more compact. This is consistent with our finding that
lines of sight toward Cep OB2 and Cep OB3 have 
many more \ion{Ca}{2} and \ion{K}{1} components. 
It is not clear, however, if the differences seen in the diffuse gas
surrounding the parent molecular clouds are 
related to the star formation histories in the clouds. 
The molecular cloud system  
associated with Cep OB3 is currently forming its second generation of 
stars, whereas a third generation of stars is forming in clouds 
associated with Cep OB2 and possibly in the $\rho$ Oph region. The older 
stellar subgroup in Cep OB3 has an age of 
$\sim$ 7 Myr, whereas the oldest stars in Cep OB2 and in Sco OB2 
(to which $\rho$ Oph belongs) are about 15 Myr, with the second older
groups $\sim$ 7 Myr (see $\S$ 2). It would be useful to 
clarify possible connections between physical conditions and differences 
in star-forming history. It seems that star formation in Cep OB2
and Cep OB3 has proceeded via sequential triggering (Patel et al. 1995;
Sargent 1979), whereas star formation scenarios for $\rho$ Oph are still
a matter of debate (de Geus 1989; Sartori et al. 2003).  
de Geus et al. (1989) derived respective ages of 11--12, 14--15, and 
5--6 Myr for three stellar subgroups, LCC, UCL, and US, of Sco OB2. 
They argued that either the 
classic picture of sequential star formation (Blaauw 1964; Elmegreen \&
Lada 1977) is not valid for the Sco OB2 Association as a whole because the 
oldest subgroup, UCL, is in between the two younger ones, or for some unknown 
reason massive star formation was initiated near the middle of the 
original giant molecular cloud. However, Pan (2002) found that uncertainties
in the above derived ages are large, and that LCC and UCL might actually 
have a similar age  and thus
star formation in Sco OB2 may not
contradict the model for sequential star formation. 
Based on spatial positions and ages of stellar subgroups in Sco OB2, 
Pan (2002) conjected that LCC and UCL could represent the first 
generation stars, the members of US the
second, and the newly formed stars in the $\rho$ Oph cloud the third
generation. If all three star-forming
regions follow the classical picture of sequential star formation 
(Elmegreen \& Lada 1977), and assuming stellar generations in Sco OB2
as that suggested by Pan (2002), then the age periods
between the 1$^{st}$ and 2$^{nd}$ and between the 2$^{nd}$ and
3$^{rd}$ are about 7--9 and 5--6 Myr in all three star-forming regions.
Support for this picture comes from 
Sartori et al. (2003). They found that the US 
subgroup is 4--8 Myr old and that LCC and UCL have the same age, 
16--20 Myr, by adopting Hipparcos distances, 
newer isochrones, and temperatures derived from the spectral types.
However, from the spatial distribution, the space velocity,  
and the age distribution of stars in Sco OB2, Sartori et al (2003) 
argued that star formation in $\rho$ Oph, most likely,
is connected to star formation in nearby spiral
arms rather than sequential triggering. This may explain why
material in $\rho$ Oph is more compact than in Cep OB2 and Cep OB2.

\section {Summary}

We obtained and analyzed 
FUV data on H$_2$ and CO along 15 lines of sight toward Cep OB2 and Cep OB3
to complement the optical survey results of Paper I. Average $b$-values of
individual components are found to be 
consistent with those obtained from ultra-high resolution spectra. The
$b$-values for CN components are also consistent with those derived from
the doublet ratio method (Str\"{o}mgren 1948; Gredel et al. 1991). 
The inferred CN excitation temperatures range from 2.55 to 2.89~K, with an 
average of 2.75 $\pm$ 0.10 K, indicating that no significant excitation
in addition to that due to the CMB
is present in the diffuse molecular clouds in our sample. 

With our large high-resolution data set, 
we examined some possible correlations between column densities  
of different species based on individual velocity components. 
In general, relationships are found to be tighter compared with those based 
on total column densities. The main correlations are the following.
(1) There are two kinds of CH in diffuse molecular gas: CN-like CH and 
CH$^+$-like CH. 
Disentangling the amount of CH in each appears possible when utilizing the 
relation between column densities of CH$^+$-like CH and of CH$^+$. 
(2) Column densities of CN and CH components are well correlated. The 
relationship becomes even stronger if the corresponding column density of 
CH$^+$-like CH is subtracted from the total CH column. A close
correspondence between CN and CO is also seen.
(3) The trends between CH and \ion{K}{1}, and \ion{Ca}{1} and \ion{K}{1} 
show tight correlatations. The slope of the relationship for 
$N$(\ion{Ca}{1}) vs. $N$(\ion{K}{1}) is 0.60 $\pm$ 0.04, significantly 
smaller than 1.0, which
may suggest that calcium depletion depends more strongly on local density 
than does potassium depletion.

We investigated the spatial distributions of species by analyzing apparent
optical depth 
profiles, distributions of components with respect to $V_{LSR}$, 
$b$-values, ionization potentials of elements, and correlations between 
column densities. These analyses show that different species are restricted 
to specific density regions. The CN and CO molecules mainly reside in 
denser regions of diffuse 
molecular clouds. No observable amount of CN is present in low density clouds 
or in cloud envelopes. The species CH and \ion{K}{1} are distributed in 
high- and moderately high-density gas ($n \ge 30$ cm$^{-3}$). (\ion{K}{1} may 
not be present in very dense regions of clouds because there may be
enhanced K depletion.) CH$^+$ and 
\ion{Ca}{1} are mainly distributed in moderately high- and intermediate-density 
regions ($n \sim$ 10--300 cm$^{-3}$). The \ion{Ca}{2} ion is 
the most widely distributed among the observed species; it can exist in 
moderately high-density regions but favors a relatively low density 
environment. 

Several correlations and analyses show that both Ca and K may be depleted onto
grains in denser gas, and that the Ca depletion has a steeper dependence on
local density. Within the framework
of Cardelli et al (1991), our data suggest that Ca column density
varies roughly as $n^{-2.9}$, whereas K column is proportional to $n^{-2.0}$
due to their depletions.
However, we note that their assumptions involving space densities and column 
densities may not be applicable. 

Gas densities for individual velocity components where CN is detected were 
inferred from a chemical model. The inferred gas densities are independent of 
assumptions about cloud shape, on which some previous calculations relied. 
Based on these derived gas densities, we examined the large- and small-scale 
structure of clouds. In general, our data are consistent with 
the large-scale structure suggested by maps of CO radio emission. 
Our analysis reveals the presence of variations in gas density among sight 
lines in two multiple star systems, $\rho$ Oph and HD~206267. The gas 
density is seen 
to vary by factors of 5\sbond 10 over scales of $\sim$ 10,000 AU. 
This indicates that observed column density variations are due in 
part to a change in
gas density. However, variations of the order of 10$^4$ suggested by 
21 cm observations (e.g., Frail et al. 1994) must have a different origin.  
Cloud thicknesses for individual components (cloudlets) are 
estimated to be $\sim$ 1pc, which may suggest that some of the cloudlets 
are sheet-like with aspect ratios of 5 to 10. The estimation of cloud thickness 
provides evidence for the need to consider non-spherical geometries 
(Heiles 1997) when analyzing small-scale structure in diffuse clouds.

Comparisons show that there are both similarities and differences in general 
characteristics of diffuse gas in the three star-forming regions, 
$\rho$ Oph, Cep OB2 and 
Cep OB3. For example, cloud turbulence in the three regions does not differ 
significantly (but note that Cep OB3 seems to have consistently larger 
$b$-values than $\rho$ Oph and Cep OB2). Clouds in Cep OB2 and Cep OB3
have more complex (clumpy) structures than those in $\rho$ Oph; in 
other words, material in $\rho$ Oph is more compact than in Cep OB2 and 
Cep OB2. The molecular cloudlets associated with Cep OB3
have lower gas density and lower total molecular mass,
compared to those associated with $\rho$ Oph and Cep OB2. 
The differences may be related to the star formation histories for the 
regions.

This paper presents analyses of the physical and chemical structure of the
diffuse gas in star-forming regions based on absorption at visible and UV 
wavelengths. As in Federman et al. (1997) and Knauth et al. (2001), such 
analyses provide results for foreground diffuse molecular gas that are
both consistent and complement results of the denser gas probed at longer
wavelength. The combined set of data yields a comprehensive view
of the effects of star formation on molecular clouds.

\acknowledgments

This research made use of the Simbad database operated at CDS 
Strasbourg, France. K. P. acknowledges KPNO for providing board and lodging 
in Tucson during observing runs. We thank the anonymous referee for 
constructive suggestions that improved the paper.
The research is based in part on observations made with the NASA--CNES--CSA
{\it Far Ultravoilet Spectroscopic Explorer} ({\it FUSE}), which is operated for
NASA by the Johns Hopkins University under NASA contract NAS5--32985. 
Additional observations made with the NASA/ESA {\it Hubble Space Telescope}
were obtained from the Multiwavelength Archive at the Space Telescope
Science Institute; STScI is operated by the Association of Universities
for Research in Astronomy, Inc. under NASA contract NAS5--26555.
This work was supported 
by NASA grants NAG5--4957, NAG5--8961, and NAG5--10305 and grants
GO--08693.03--A and AR--09921.01--A from the Space Telescope Science 
Institute. We acknowledge
use of the regression subroutine obtained from the Penn State statistical
software archive.

\clearpage

\begin{deluxetable}{lllrl}
\tabletypesize{\footnotesize}
\tablewidth{0pt}
\tablecaption{FUV Data Summary}
\tablecolumns{5}
\tablehead{
\colhead{Star}
&\colhead{Satellite}
&\colhead{Spectra}
&\colhead{S/N}
&\colhead{Molecule}
}
\startdata
\multicolumn{5}{c}{Cep OB2}\\
\hline
HD203374A& {\it FUSE}& B03001& 50& H$_2$\\
         & {\it HST} & O5LH08& 25& CO   \\
HD206165/9 Cep & {\it FUSE}& B03002& 25& H$_2$, CO\\
HD206267A& {\it HST} & O5LH09& 30& CO   \\
HD206773 & {\it FUSE}& B07109& 70& H$_2$, CO\\
         & {\it HST} & O5C04T& 25& CO   \\
HD207198 & {\it HST} & O59S06& 25& CO   \\
HD207308 & {\it FUSE}& B03003& 20& H$_2$\\
         & {\it HST} & O63Y02& 60& CO   \\
HD207538 & {\it HST} & O63Y01& 55& CO   \\
HD208266 & {\it HST} & O63Y03& 60& CO   \\
HD208440 & {\it FUSE}& B03004& 75& H$_2$\\
         & {\it HST} & O5C06M& 25& CO   \\
HD209339 & {\it FUSE}& B03005& 70& H$_2$\\
         & {\it HST} & O5LH0B& 35& CO   \\
HD210839/$\lambda$ Cep & {\it HST} & O54304& 45& CO   \\
\hline
\multicolumn{5}{c}{Cep OB3}\\
\hline
HD216532 & {\it FUSE}& A05102& 15& H$_2$, CO\\
HD216898 & {\it FUSE}& A05103& 20& H$_2$, CO\\
HD217035A& {\it FUSE}& A05104& 25& H$_2$, CO\\
HD217312 & {\it FUSE}& P19305& 35& H$_2$, CO
\enddata
\end{deluxetable}

\begin{deluxetable}{lcccccc}
\tabletypesize{\footnotesize}
\tablewidth{0pt}
\tablecaption{Molecular Results\tablenotemark{a}}
\tablehead{
\colhead{Target}
&\colhead{$N$(H$_2$)}
&\colhead{$T_{\rm 1,0}$}
&\colhead{$N$(CH)}
&\colhead{$N$(CO)}
&\colhead{CH/H$_2$}
&\colhead{CO/H$_2$}
\\
\colhead{}
&\colhead{(10$^{20}$ cm$^{-2}$)}
&\colhead{(K)}
&\colhead{(10$^{13}$ cm$^{-2}$)}
&\colhead{(10$^{15}$ cm$^{-2}$)}
&\colhead{(10$^{-8}$)}
&\colhead{(10$^{-6}$)}
}
\startdata
\multicolumn{7}{c}{Cep OB2}\\
\hline
HD203374A& 5.04(0.08)& 76(1)& 2.22(0.2)& 2.56(0.11)& 4.40(0.45)& 5.08(0.23) \\
9 Cep & 5.98(0.26)& 68(4)& 2.04(0.2)& 1.53(0.04)& 3.41(0.37)& 2.56(0.13) \\
HD206267A& 7.24(0.63)& 65(5)& 3.03(0.3)& 10.1(0.7) & 4.19(0.56)& 14.0(1.6) \\
HD206773 & 2.93(0.05)& 94(1)& 1.12(0.1)& 0.23(0.03)& 3.82(0.39)& 0.80(0.09) \\
HD207198 & 6.76(0.59)& 66(5)& 3.53(0.4)& 2.65(0.12)& 5.22(0.69)& 3.92(0.39) \\
HD207308 & 7.28(0.17)& 57(1)& 3.21(0.3)& 8.60(0.95)& 4.41(0.45)& 11.8(1.3) \\
HD207538 & 8.13(1.05)& 73(8)& 3.81(0.4)& 2.49(0.25)& 4.69(0.77)& 3.06(0.50) \\
HD208266 &7.4--10\tablenotemark{b}&\nodata& 3.35(0.3) & 11.2(1.2)
& \nodata   & \nodata \\
HD208440 & 2.21(0.03)& 75(2)& 1.16(0.1)& 0.16(0.01)& 5.25(0.53)& 0.72(0.03) \\
HD209339 & 1.79(0.03)& 90(1)& 0.79(0.1)& 0.09(0.01)& 4.41(0.45)& 0.49(0.06) \\
$\lambda$ Cep & 6.92(0.61)& 72(6)& 2.08(0.2)& 2.47(0.20)& 3.01(0.40)& 3.57(0.43) \\
\hline
\multicolumn{7}{c}{Cep OB3}\\
\hline
HD216532 & 12.7(0.4) & 83(8)& 2.02(0.2)& 1.42(0.11)& 1.59(0.17)& 1.12(0.09) \\
HD216898 & 11.1(0.3) & 86(6)& 2.85(0.3)& 1.09(0.08)& 2.12(0.22)& 0.98(0.08) \\
HD217035A& 9.00(0.25)& 75(1)& 1.68(0.2)& 0.37(0.03)& 1.87(0.19)& 0.41(0.03) \\
HD217312 & 6.38(0.16)& 76(3)& 2.56(0.3)& 0.24(0.03)& 4.01(0.41)& 0.38(0.05) \\
\enddata
\tablenotetext{a}{Uncertainties are denoted within parentheses.}
\tablenotetext{b}{$N$(H$_2$) is predicted from $N$(CH) and $N$(CO).} 
\end{deluxetable}

\begin{deluxetable}{lcccccccc}
\tablecolumns{9}
\tabletypesize{\scriptsize}
\tablewidth{0pt}
\tablecaption{Column Densities for H$_2$ Rotational Levels\tablenotemark{a}}
\tablehead{
\colhead{Target}
&\multicolumn{8}{c}{Log$_{10}$($N_J$)}
\\
\colhead{}
&\colhead{Total} &\colhead{$J$=0}
&\colhead{$J$=1} &\colhead{$J$=2}
&\colhead{$J$=3} &\colhead{$J$=4}
&\colhead{$J$=5} &\colhead{$J$=6}
}
\startdata
\multicolumn{9}{c}{Cep OB2}\\
\hline
HD203374A& 20.70(0.01)& 20.41(0.01)& 20.39(0.02)& 18.59(0.12)& 18.17(0.06)
& 15.91(0.04) & 14.95(0.14)& 13.94(0.12) \\
9 Cep & 20.78(0.02)& 20.53(0.01)& 20.40(0.06)& 18.74(0.08)& 18.21(0.15)
& 17.64(0.20) & 15.80(0.41)& 14.97(0.05) \\
HD207308 & 20.86(0.01)& 20.70(0.01)& 20.35(0.03)& 18.70(0.10)& 17.86(0.12)
& 17.14(0.14) & 16.00(0.37)& 14.15(0.12) \\
HD208440 & 20.34(0.01)& 20.05(0.01)& 20.02(0.02)& 18.42(0.02)& 18.13(0.11)
& 16.04(0.17) & 15.02(0.11)& 14.18(0.16) \\
HD209339 & 20.25(0.01)& 19.87(0.01)& 20.00(0.01)& 18.36(0.05)& 18.02(0.07)
& 17.41(0.05) & 16.40(0.07)& 14.06(0.04) \\
\hline
\multicolumn{9}{c}{Cep OB3}\\
\hline
HD216532 & 21.10(0.02)& 20.76(0.04)& 20.83(0.06)& 19.27(0.05)& 17.93(0.31)
& 17.52(0.30) & 17.03(0.12)& 14.59(0.18) \\
HD216898 & 21.05(0.02)& 20.69(0.02)& 20.78(0.04)& 19.08(0.07)& 18.31(0.08)
& 17.19(0.20) & 15.37(0.15)& 14.37(0.09) \\
HD217035A& 20.95(0.02)& 20.66(0.01)& 20.63(0.02)& 18.98(0.08)& 18.34(0.23)
& 17.91(0.27) & 15.97(0.19)& 14.49(0.19) \\
\enddata

\tablenotetext{a}{Uncertainties are denoted within parentheses.}
%\tablenotetext{b}{From our {\it FUSE} programs B030 (Cep OB2) and A051 
%(Cep OB3).}
\end{deluxetable}

\begin{deluxetable}{lccccccc}
\tablecaption{Average $b$-values (in km s$^{-1}$) for Individual Components}
\tablehead{ & &\multicolumn{3}{c}{Star-forming Region} & & 
\multicolumn{2}{c}{Overall} \\ \cline{3-5} \cline{7-8} 
Species & & Cep OB2 & Cep OB3 & $\rho$ Oph & & $b$ & \# of components }
\startdata
CN && 0.83$\pm$0.11 & 0.84$\pm$0.11 & 0.80$\pm$0.05&& 0.83$\pm$0.11 & 50\\
CO && 0.59$\pm$0.20 & 0.77$\pm$0.35 & \nodata && 0.66$\pm$0.28 & 53\\
%CO && 0.66$\pm$0.30 & 0.82$\pm$0.35 & $\ldots$     && 0.69$\pm$0.32 & 125\\
CH && 1.04$\pm$0.17 & 1.10$\pm$0.19 & 0.91$\pm$0.11&& 1.04$\pm$0.18 & 125\\
CN-like CH && 0.90$\pm$0.11 & $\ldots$ & $\ldots$&& 0.90$\pm$0.11 & 12\\
CH$^+$-like CH && 1.10$\pm$0.15 & 1.12$\pm$0.18 & 0.96$\pm$0.14&& 
1.10$\pm$0.16 & 78\\
CH$^+$ && 1.96$\pm$0.23 & 1.97$\pm$0.23 & 1.80$\pm$0.23&& 1.96$\pm$0.23& 111\\
\ion{Ca}{1}&& 0.94$\pm$0.16 & 1.07$\pm$0.13 & 1.03$\pm$0.11&& 
0.96$\pm$0.16 & 119\\
\ion{K}{1}&& 0.88$\pm$0.21 & 1.08$\pm$0.22 & 0.87$\pm$0.16&& 
0.92$\pm$0.22 & 224 \\
\ion{Ca}{2}&& 1.61$\pm$0.26 & 1.75$\pm$0.19 & 1.35$\pm$0.40&& 
1.62$\pm$0.28 & 350 \\
\enddata

\end{deluxetable}

\begin{deluxetable}{lcccccccccc}
\tabletypesize{\scriptsize}
\tablewidth{0pt}
\tablecaption{Molecular Correlations for Cloud Components toward Cep OB2}
\tablehead{
\colhead{Target}
&\colhead{Species}
&\multicolumn{7}{c}{Cloud Component Structure}
&\colhead{RMS\tablenotemark{a}}
&\colhead{\case{\rm RMS(CH)}{\rm RMS(CN)}}\\
\cline{3-9}
}
\startdata
HD203374A& CO& .01& .00& .50& .14& .34& .00& .01&\nodata& \nodata\\
         & CN& \nodata& \nodata& .46& .18& .36& \nodata& \nodata& 0.0233& \nodata \\
         & CH& .05& .04& .32& .20& .28& .06& .05& 0.0828& 3.6 \\
\hline
HD206267A& CO& .07& .83& .07& .02& .00& .01&    &\nodata& \nodata\\
         & CN& .03& .81& .10& .06& \nodata& \nodata&    & 0.0277& \nodata \\
         & CH& .14& .29& .07& .29& .07& .14&    & 0.2553& 9.2 \\
\hline
HD206773 & CO& .18& .72& .10&    &    &    &    &\nodata& \nodata\\
         & CN& \nodata&1.00& \nodata&    &    &    &    & 0.2007& \nodata \\
         & CH& .14& .40& .46&    &    &    &    & 0.2790& 1.4 \\
\hline
HD207198 & CO& .22& .09& .47& .16& .02& .04&    &\nodata& \nodata\\
         & CN& .28& \nodata& .52& .20& \nodata& \nodata&    & 0.0545& \nodata \\
         & CH& .15& .08& .29& .28& .08& .12&    & 0.1015& 1.9 \\
\hline
HD207308 & CO& .12& .00& .84& .04&    &    &    &\nodata& \nodata\\
         & CN& \nodata& \nodata& .82& .18&    &    &    & 0.0927& \nodata \\
         & CH& .07& .09& .57& .27&    &    &    & 0.1847& 2.0 \\
\hline
HD207538 & CO& .01& .05& .33& .59& .02& .00&    &\nodata& \nodata\\
         & CN& \nodata& .12& .31& .38& .19& \nodata&    & 0.1143& \nodata \\
         & CH& .06& .18& .31& .31& .11& .03&    & 0.1337& 1.2 \\
\hline
HD208266 & CO& .00& .87& .12& .00& .01&    &    &\nodata& \nodata\\
         & CN& \nodata& .77& .16& .07& \nodata&    &    & 0.0576& \nodata \\
         & CH& .01& .39& .37& .12& .11&    &    & 0.2520& 4.4 \\
\hline
$\lambda$ Cep & CO& .72& .28&    &    &    &    &    &\nodata& \nodata\\
         & CN& .74& .26&    &    &    &    &    & 0.0200& \nodata\\
         & CH& .62& .38&    &    &    &    &    & 0.1000& 5.0
\enddata

\tablenotetext{a}{Of the CO structure $vs.$ the CN or CH structure. Undetected
CN components (\nodata) were assigned the value ``.00'' for rms calculations.
CO components designated by ``.00'' were modeled to be $\leq$ .005.}
\end{deluxetable}

\clearpage

\begin{deluxetable}{lcccccccc}
\tablewidth{0pt}
%\tablecolumns{10}
\tablecaption{Relationships among Column Densities Involving Individual 
Components} 
\tablehead{
   &     &       &   &  &&\multicolumn{3}{c}
   {log[$N$(Y)] = A + B $\times$ log[$N$(X)]} \\  \cline{7-9} 
 Y   & X   & $N$  & $r$ & $p$ && A  & B  & d$_{rms}$ \tablenotemark{a} } 
\startdata 
CH   &  CH$^+$ & 90&0.37& 0.0004 &&5.65$\pm$1.30 & 0.55$\pm$0.11 & 0.288 \\
CH$^+$-like CH \tablenotemark{b}& CH$^+$ & 36 & 0.77 &0.0000 && 0.60$\pm$1.34 
& 0.95$\pm$0.13 &0.124 \\
CN & CH & 48 &0.80 &0.0000 && $-$12.54$\pm$1.98 & 1.91$\pm$0.16 & 0.127  \\
CN & CN-like CH &12& 0.97&0.0000 && $-$10.25$\pm$ 1.72 & 1.74$\pm$0.13 
& 0.053\\
CN & CH$^+$-cor CH & 40 & 0.87 & 0.0000 && $-$10.08$\pm$1.76 & 
1.73$\pm$0.13 & 0.120 \\
\ion{Ca}{1} & \ion{Ca}{2} &117& 0.47& 0.0000 && $-$0.08$\pm$0.96 
& 0.81$\pm$0.08 &0.220  \\
\ion{Ca}{1} & \ion{K}{1} & 114 & 0.69 &0.0000 && 2.88$\pm$0.47 & 
0.60$\pm$0.04 & 0.171 \\
\ion{K}{1} &CH& 123 &0.87 &0.0000 && $-$0.85$\pm$0.55& 0.96$\pm$0.04 & 0.134\\
\ion{K}{1} &\ion{Ca}{2} & 223 & 0.55 & 0.0000 && $-$10.94$\pm$0.72 & 
1.89$\pm$0.06 & 0.277 \\
%CN/CH & \ion{Ca}{1}/CH & 37 & $-$0.68 & 0.0000&&$-$4.03$\pm$0.46 & 
%$-$1.03$\pm$0.15 & 0.139 \\
%CN/CH & \ion{K}{1}/CH & 48 & $-$0.40 & 0.005 &&$-$3.30$\pm$0.34 & 
%$-$2.03 $\pm$ 0.20 & 0.121 \\
\enddata

\tablenotetext{a}{Root mean square distance of points from the fit line.}
\tablenotetext{b}{Only components with log[$N$(CH$^+)]\le $ 12.6 are fitted.}      
\end{deluxetable}

\clearpage

\begin{deluxetable}{lccccrrrrr}
%\rotate
\tablecolumns{10}
\tablewidth{0pt}
\tabletypesize{\scriptsize}
\tablecaption{Chemical Results}
\tablehead{
Cloud\tablenotemark{a} & $I_{uv}$ & $\tau_{uv}$ & $T$ & $n$ & 
$N_o$(CH) & $N_o$(C$_2$) & $N_p$(C$_2$) & $N_o$(CN) & 
$N_p$(CN) \\ \cline{6-10}
 & & & (K) & (cm$^{-3}$) & \multicolumn{5}{c}{($10^{12}$ cm$^{-2}$)} }
\startdata
\multicolumn{10}{c}{$\rho$ Oph} \\
$\rho$~Oph~D & 1 & 2.08 & 50 & 425 & 19.5 & $\ldots$ & 18.7 & 2.1 & 
2.1 \\
$\rho$~Oph~C & 1 & 2.04 & 40 & 1100 & 19.2 & $\ldots$ & 44.4 & 6.0 & 
6.0 \\
$\rho$~Oph~A & 1 & 2.04 & 50 & 625 & 14.9 & 26.0 & 21.4 & 2.1 & 2.6 \\
$\rho$~Oph~B & 1 & 2.04 & 50 & 450 & 16.8 & $\ldots$ & 18.0 & 2.0 & 
2.0 \\ \hline
\multicolumn{10}{c}{Cep OB2} \\
HD~203374A($-1.2$) & 1 & 3.69 & 50 & 90 & 6.9 & $\ldots$ & 8.5 & 1.0 
& 0.9 \\
HD~203374A($+1.2$) & 1 & 3.69 & 50 & 60 & 4.5 & $\ldots$ & 3.8 & 0.4 
& 0.4 \\
HD~203374A($+3.1$) & 1 & 3.69 & 50 & 80 & 6.3 & $\ldots$ & 7.0 & 0.8 
& 0.8 \\
HD~204827($-9.2$) & 1 & 6.66 & 50 & 30 & 7.0 & $\ldots$ & 4.0 & 1.3 
& 1.3 \\
HD~204827($-6.4$) & 1 & 6.66 & 50 & 30 & 5.3 & $\ldots$ & 3.1 & 0.9 
& 1.0 \\
HD~204827($-4.3$) & 1 & 6.66 & 30 & 200 & 31.4 & $\ldots$ & 45.3 & 
17.1 & 17.5 \\
HD~204827($-2.0$) & 1 & 6.66 & 30 & 110 & 13.6 & $\ldots$ & 16.6 & 
5.8 & 5.8 \\
HD~204827($+0.5$) & 1 & 6.66 & 30 & 125 & 26.9 & $\ldots$ & 34.2 & 
12.3 & 12.3 \\
9~Cep($-1.3$) & 1 & 2.72 & 50 & 200\tablenotemark{b} & 5.0 & 
3.5\tablenotemark{c} & 5.3 & 0.7 & 0.5 \\
9~Cep($+0.3$) & 1 & 2.72 & 50 & 300\tablenotemark{b} & 5.7 & 
6.0\tablenotemark{c} & 8.7 & 1.2 & 0.9 \\
9~Cep($+3.7$) & 1 & 2.72 & 50 & 175\tablenotemark{b} & 4.1 & 
2.5\tablenotemark{c} & 3.8 & 0.5 & 0.4 \\
HD~206183 & 1 & 2.56 & 50 & 250 & 7.8 & $\ldots$ & 8.7 & 1.0 & 1.0 \\
HD~206267A($-4.7$) & 1 & 3.13& 50 & 80 & 4.1 & 2.6\tablenotemark{c} & 
2.7 & 0.3 & 0.3 \\
HD~206267A($-2.7$) & 1 & 3.13 & 50 & 1150\tablenotemark{b} & 8.8 & 
61.1\tablenotemark{c} & 49.8 & 7.0 & 10.1 \\
HD~206267A($-0.9$) & 1 & 3.13 & 50 & 525\tablenotemark{b} & 2.1 & 
7.9\tablenotemark{c} & 7.1 & 0.9 & 1.1 \\
HD~206267A($+0.9$) & 1 & 3.13 & 50 & 60 & 9.0 & 4.4\tablenotemark{c} & 
4.5 & 0.5 & 0.4 \\
HD~206267C($-3.0$) & 1 & 2.82 & 50 & 725 & 9.8 & $\ldots$ & 33.4 & 
4.9 & 4.9 \\
HD~206267C($-0.8$) & 1 & 2.82 & 50 & 250 & 2.4 & $\ldots$ & 3.4 & 
0.4 & 0.4 \\
HD~206267C($+0.8$) & 1 & 2.82 & 50 & 425 & 8.9 & $\ldots$ & 19.9 & 
2.6 & 2.5 \\
HD~206267D($-4.3$) & 1 & 2.26 & 50 & 400 & 8.0 & $\ldots$ & 10.4 & 
1.2 & 1.2 \\
HD~206267D($-0.8$) & 1 & 2.26 & 50 & 650 & 5.3 & $\ldots$ & 10.6 & 
1.3 & 1.3 \\
HD~206773 & 1 & 2.50 & 50 & 225 & 4.5 & $\ldots$ & 4.3 & 0.45 & 0.46 \\
HD~207198($-6.2$) & 1 & 3.63 & 50 & 150 & 5.3 & 8.5\tablenotemark{c} & 
9.8 & 1.3 & 1.1 \\
HD~207198($-2.1$) & 1 & 3.63 & 50 & 150 & 10.1 & 15.6\tablenotemark{c} & 
18.6 & 2.4 & 2.2 \\
HD~207198($-0.2$) & 1 & 3.63 & 50 & 50 & 9.8 & 5.9\tablenotemark{c} & 
6.6 & 0.9 & 0.7 \\
$\nu$~Cep & 1 & 2.10 & 50 & 300 & 5.1 & $\ldots$ & 4.4 & 0.45 & 0.46 \\
HD~207308($-2.4$) & 1 & 3.00 & 50 & 600 & 18.4 & $\ldots$ & 62.3 & 
9.0 & 9.1 \\
HD~207308($-0.5$) & 1 & 3.00 & 50 & 300 & 8.6 & $\ldots$ & 16.7 & 
2.0 & 2.0 \\
HD~207538($-4.4$) & 1 & 3.92 & 50 & 60 & 7.0 & $\ldots$ & 5.0 & 
0.6 & 0.6 \\
HD~207538($-2.5$) & 1 & 3.92 & 50 & 90 & 11.7 & $\ldots$ & 12.0 & 
1.5 & 1.5 \\
HD~207538($-0.5$) & 1 & 3.92 & 50 & 110 & 11.8 & $\ldots$ & 14.5 & 
1.8 & 1.8 \\
HD~207538($+1.9$) & 1 & 3.92 & 50 & 150 & 4.4 & $\ldots$ & 7.0 & 
0.9 & 0.9 \\
 & 5 & 3.92 & 50 & 700 & 4.4 & $\ldots$ & 6.6 & 0.9 & 0.9 \\
 & 1 & 3.92 & 50 & 275 & 2.3\tablenotemark{d} & $\ldots$ & 5.8 & 
0.9 & 0.9 \\
HD~208266($-4.4$) & 1 & 3.10 & 50 & 375 & 13.0 & $\ldots$ & 33.0 & 
4.4 & 4.4 \\
HD~208266($-2.8$) & 1 & 3.10 & 50 & 90 & 12.4 & $\ldots$ & 8.9 & 
0.9 & 0.9 \\
HD~208266($-0.6$) & 1 & 3.10 & 50 & 125 & 4.0 & $\ldots$ & 3.9 & 
0.4 & 0.4 \\
13~Cep($-2.0$) & 1 & 4.48 & 50 & 325 & 17.4 & $\ldots$ & 34.0 & 
7.5 & 7.7 \\
13~Cep($-0.3$) & 1 & 4.48 & 50 & 125 & 9.1 & $\ldots$ & 9.5 & 
1.7 & 1.7 \\
13~Cep($+1.5$) & 1 & 4.48 & 50 & 150 & 19.4 & $\ldots$ & 23.1 & 
4.5 & 4.3 \\
14~Cep & 1 & 1.37 & 50 & 725 & 4.2 & $\ldots$ & 4.2 & 0.45 & 0.45 \\
$\lambda$~Cep($-2.3$) & 1 & 2.36 & 50 & 425 & 12.9\tablenotemark{e} & 
12.6\tablenotemark{c} & 19.4 & 2.6 & 2.1 \\
$\lambda$~Cep($-0.4$) & 1 & 2.36 & 50 & 225 & 7.9\tablenotemark{e} & 
4.4\tablenotemark{c} & 6.6 & 0.9 & 0.7 \\ \hline
\multicolumn{10}{c}{Cep OB3} \\
HD~216532 & 1 & 2.50 & 65 & 550 & 11.2 & $\ldots$ & 23.3 & 3.1 & 
3.0 \\
HD~216898 & 1 & 3.59 & 65 & 100 & 5.4 & $\ldots$ & 6.6 & 0.8 & 0.8 \\
HD~217312($+0.0$) & 1 & 2.86 & 65 & 350 & 1.6 & $\ldots$ & 3.1 & 
0.4 & 0.4 \\
HD~217312($+5.2$) & 1 & 2.86 & 65 & 175 & 4.2 & $\ldots$ & 4.4 & 
0.5 & 0.5 \\
HD~217312($+7.2$) & 1 & 2.86 & 65 & 110 & 8.1 & $\ldots$ & 5.5 & 
0.6 & 0.6 \\
\enddata
\tablenotetext{a}{If more than one cloud containing CN appears along a 
line of sight, the velocity is given in parentheses.}
\tablenotetext{b}{Average of results for C$_2$ and CN.}
\tablenotetext{c}{Scaled line-of-sight $N$(C$_2$) to relative 
contributions from CN components.}
\tablenotetext{d}{Column associated with CN-like CH only.}
\tablenotetext{e}{$N$(CH) from Crane et al. 1995.}
\end{deluxetable}

\begin{deluxetable}{lccccccc}
\tablecaption{Mean Number of Velocity Components per Sight Line and Average 
Column Density per Component}
\tablehead{& \multicolumn {3}{c}{$\langle$ M $\rangle$ } & &\multicolumn {3}
{c}{$\langle N \rangle$ (cm$^{-2}$)}\\  \cline{2-4} \cline{6-8}
Species & Cep OB2 & Cep OB3 & $\rho$ Oph & &Cep OB2 & Cep OB3 & $\rho$ Oph }
\startdata 
\ion{Ca}{2} & 13.0 & 13.6 & 5.5 && 4.43 $\times~10^{11}$ & 4.84 
$\times~10^{11}$ & 3.44 $\times~10^{11}$\\ 
\ion{K}{1} & 8.0 & 9.8 & 4.0 && 2.12 $\times~10^{11}$ & 1.52 
$\times~10^{11}$ & 2.51 $\times~10^{11}$\\ 
\ion{Ca}{1} & 4.7 & 4.5 & 2.0 && 5.40 $\times~10^{9}$ & 3.84 
$\times~10^{9}$ & 8.83 $\times~10^{9}$\\ 
CH$^+$      & 3.9 & 6.6 & 2.0 && 3.56 $\times~10^{12}$ & 5.24 
$\times~10^{12}$ & 5.58 $\times~10^{12}$\\ 
CH          & 4.5 & 5.8 & 2.2 && 5.70 $\times~10^{12}$ & 3.53 
$\times~10^{12}$ & 9.16 $\times~10^{12}$\\ 
CN          & 2.0 & 1.0 & 1.0 && 2.60 $\times~10^{12}$ & 1.08 
$\times~10^{12}$ & 3.05 $\times~10^{12}$\\ 
\enddata

\end{deluxetable}

\clearpage

\begin{figure}
%\plotone{cnchn.ps}
\plotone{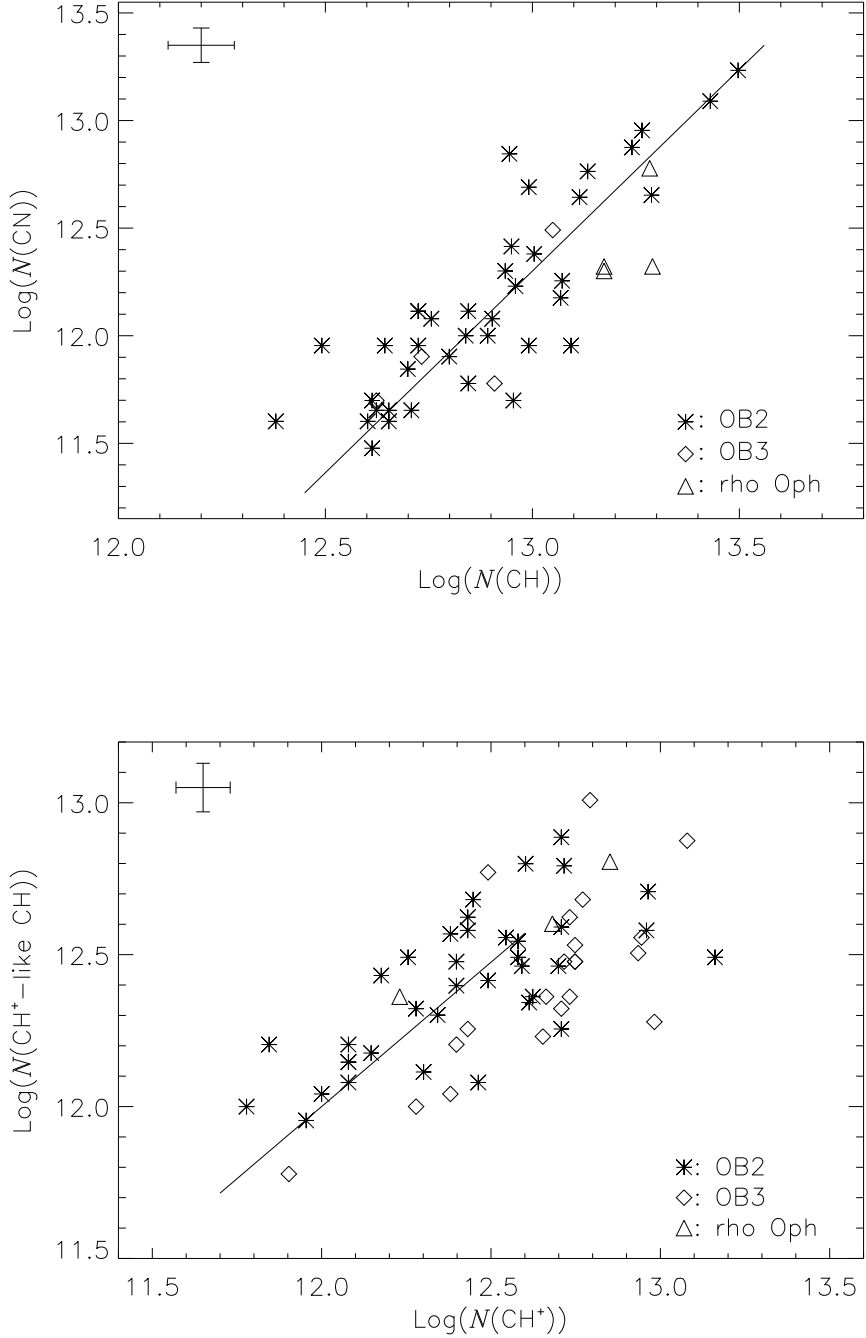}
\caption{Logarithmic plots of column densities. The error bars represent 20\% 
uncertainties in $N$.~~$Top$: $N$(CN) vs. $N$(CH). The solid line presents our
best fit
\newline ~~$Bottom$: $N$(CH$^+$-like~CH) vs. $N$(CH$^+$). The solid line
is the best fit to components with log[$N$(CH$^+$)] $\le$ 12.6 (see text for
more detail). Here, and elsewhere, components toward Cep OB2 (asterisks), 
Cep OB3 (diamonds), and $\rho$ Oph (triangles) are distinguished. }
\end{figure}

\clearpage

\begin{figure}
%\plotone{cnchcn.ps}
\plotone{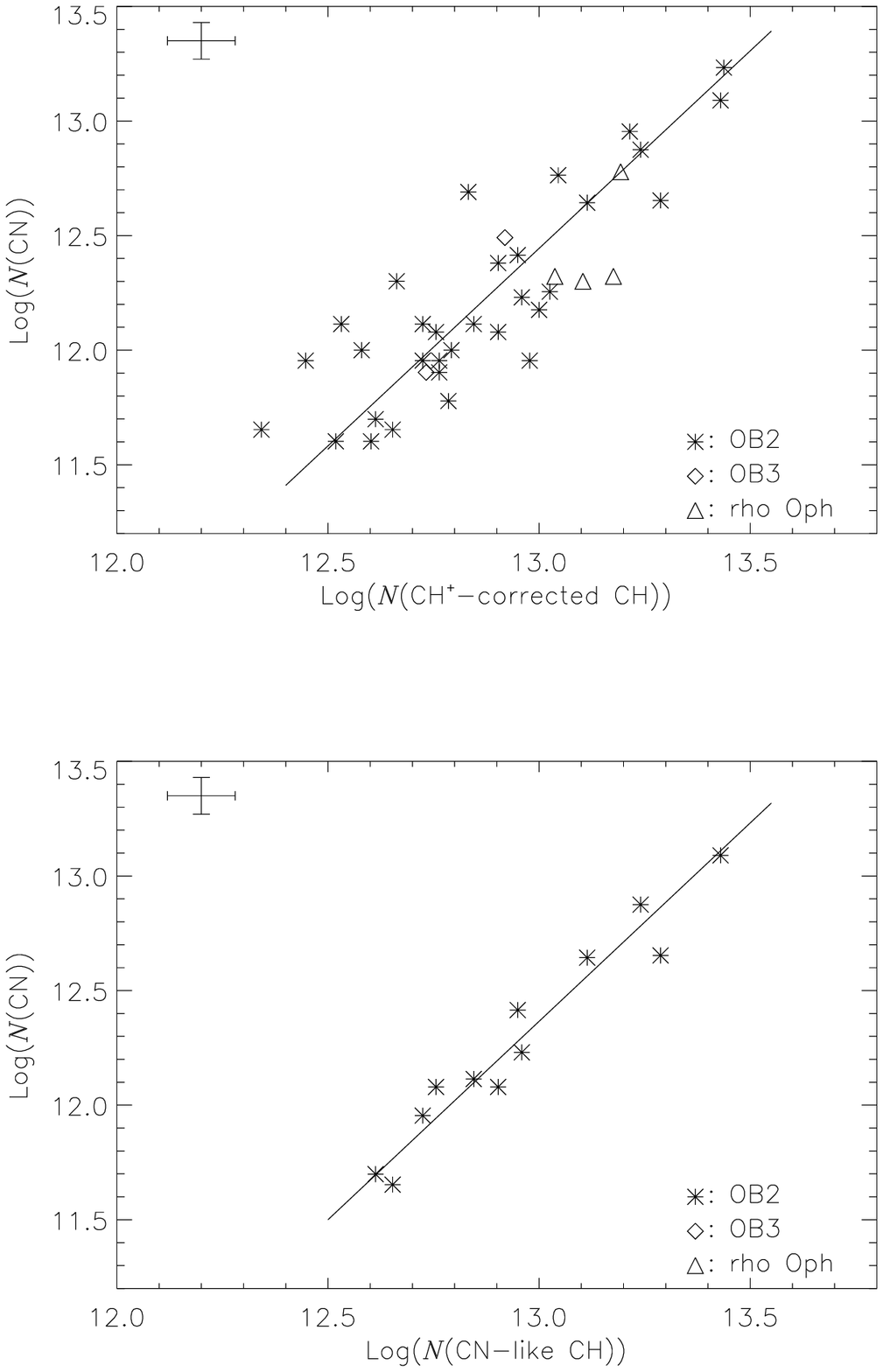}
\caption{Logarithmic plots of column densities. The error bars represent 20\% 
uncertainties in $N$.~~$Top$: $N$(CN) vs. $N$(CH$^+$-corrected CH). The solid 
line presents the best fit with a slope of 1.73, indistinguishable from
the one between $N$(CN) and $N$(CN-like CH).~~$Bottom$: $N$(CN) vs. 
$N$(CN-like CH).}
\end{figure}

\clearpage

\begin{figure}
%\plotone{chkn.ps}
\plotone{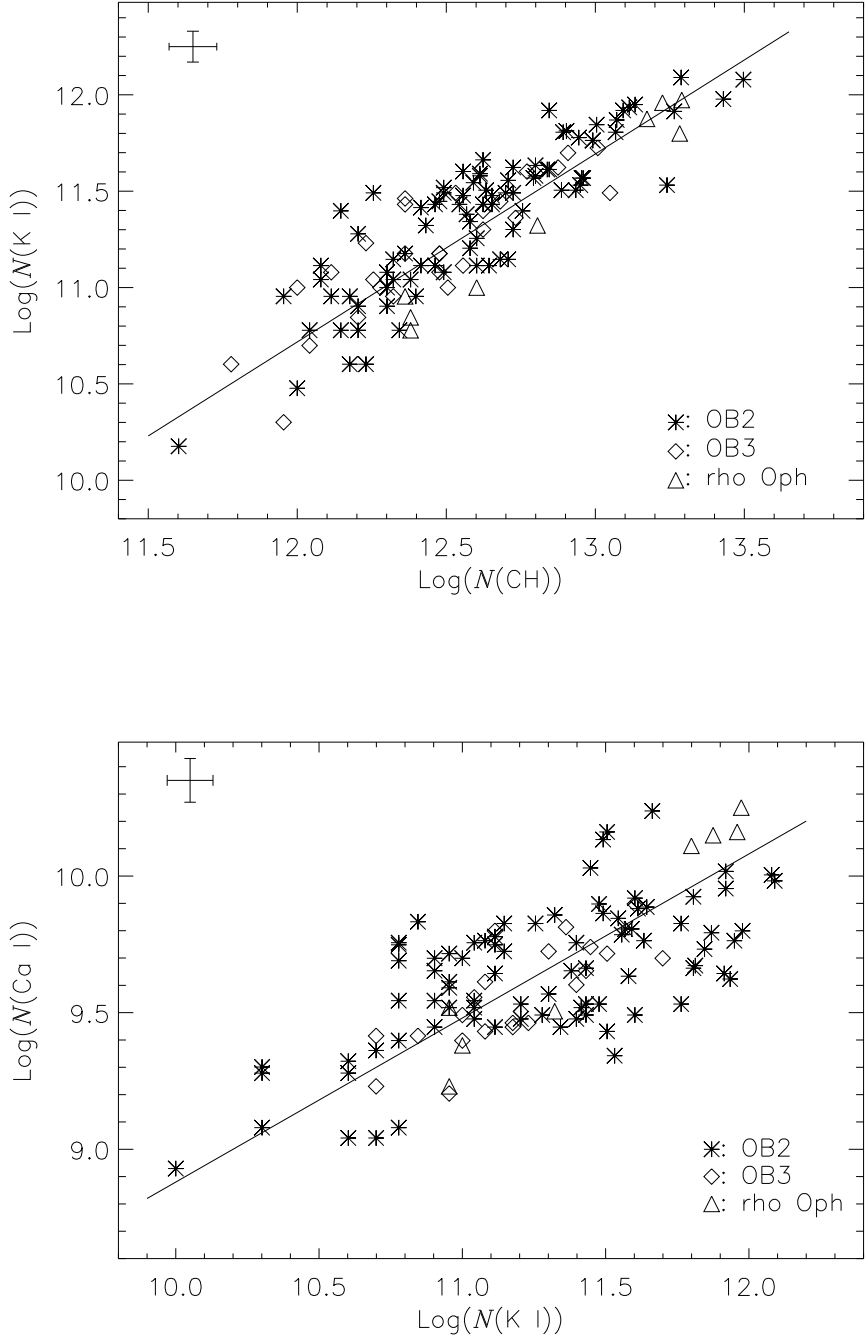}
\caption{Logarithmic plots of column densities. The error bars represent 20\% 
uncertainties in $N$.~~$Top$: $N$(\ion{K}{1}) vs. $N$(CH). ~~$Bottom$: 
$N$(\ion{Ca}{1}) vs. $N$(\ion{K}{1}). The solid 
line presents the best fit with a slope of 0.60 $\pm$ 0.04, which may 
suggest increasing calcium depletion in denser gas.}
\end{figure}

\clearpage

\begin{figure}
%\epsscale{0.9}
%\plotone{hd207308_optdepth.ps}
\plotone{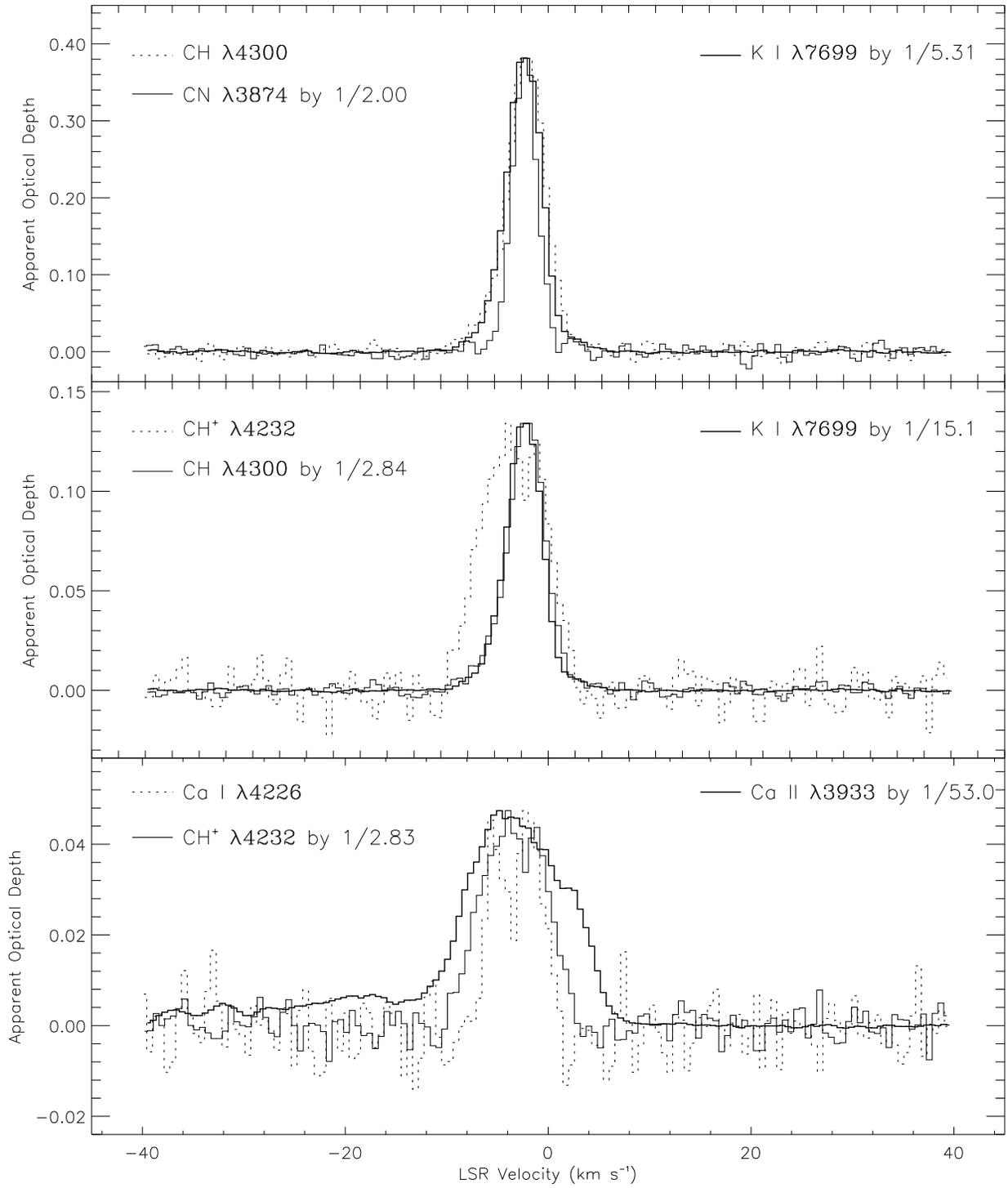}
\vspace{0.06in}
\caption{Plots of apparent optical depth profiles for different species on
the line of sight toward HD~207308. Scaling factors are applied to individual
profiles for easier comparison.}
\end{figure}

\clearpage

\begin{figure}
%\epsscale{0.9}
%\plotone{numberdis.ps}
\plotone{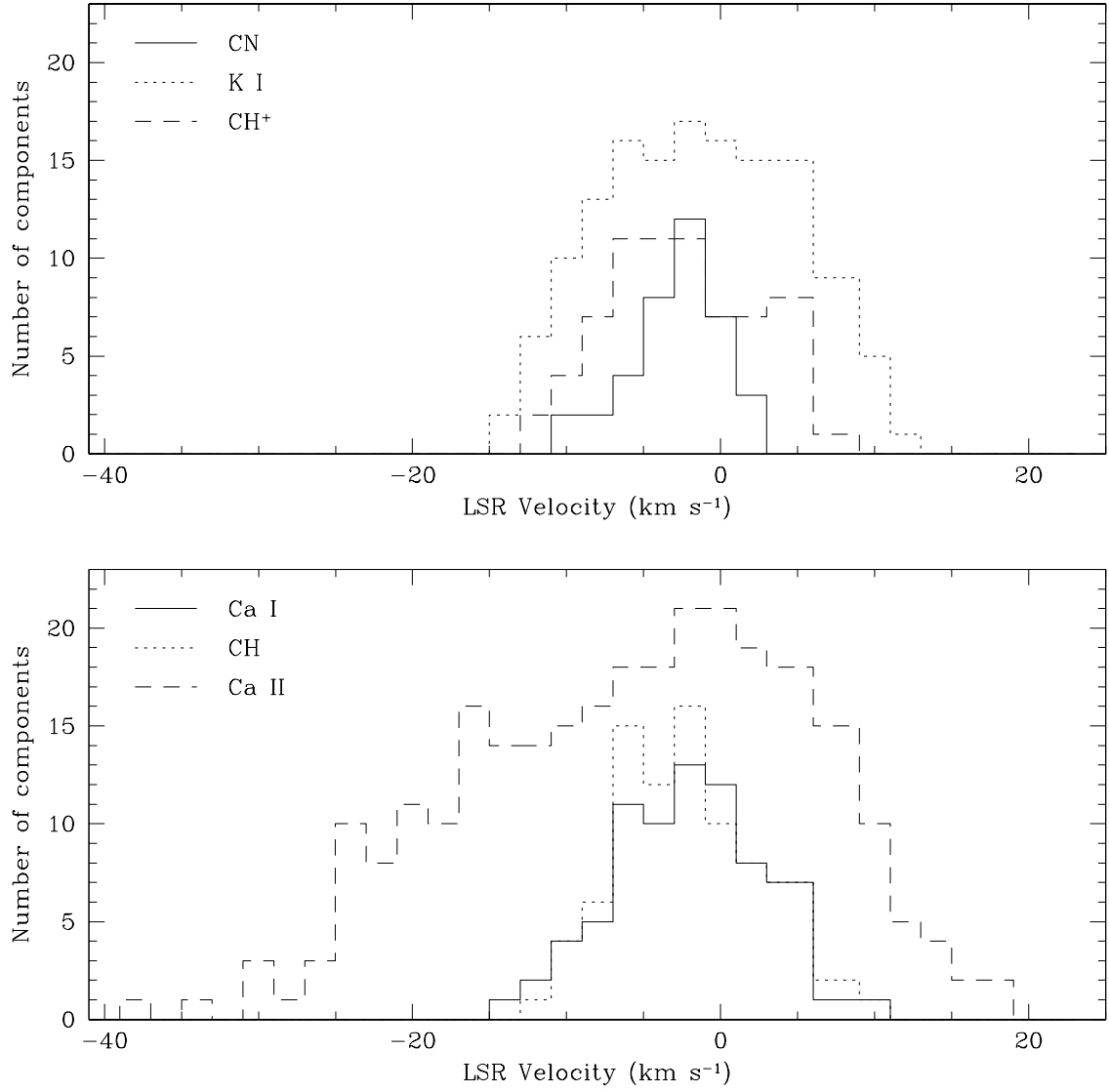}
\vspace{0.06in}
\caption{Distributions of velocity components along lines of sight toward 
Cep~OB2 with respect to $V_{LSR}$.}
\end{figure}

\clearpage

\begin{figure}
\vspace{-1.00in}
%\plotone{schematic.eps}
\plotone{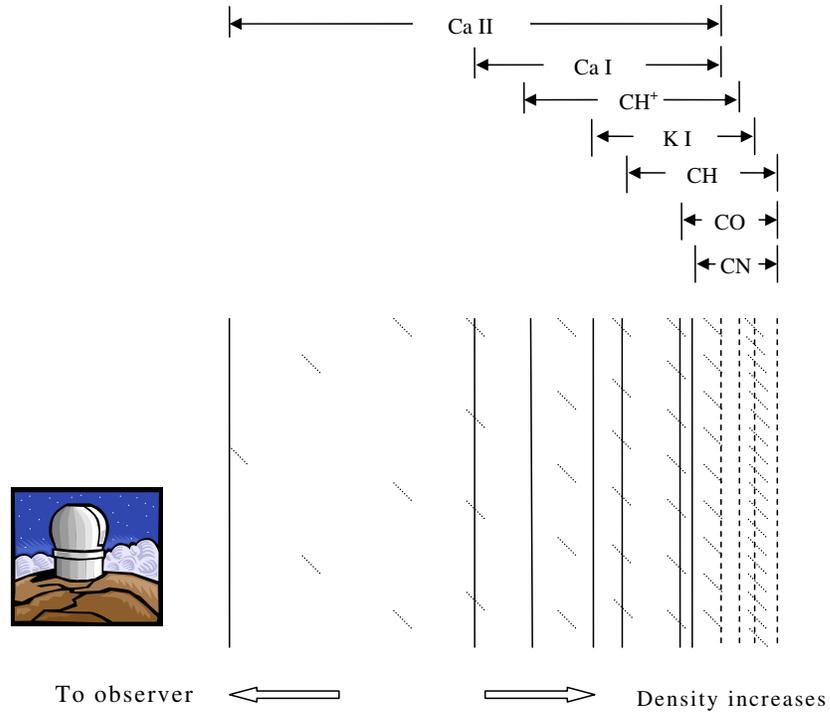}
\vspace{-2.50in}
\caption{A schematic showing the inferred distributions of species in a 
  diffuse cloud, here represented by a plane---parallel slab from the
  cloud surface to its denser regions. Different species are distributed 
  according to gas density in the diffuse molecular gas (see text in 
  $\S$7.1). Dashed and solid vertical lines indicate the inner and outer
  boundaries of species distributions. In some low density clouds, CN and CH 
  are not present. If the density is even lower, \ion{Ca}{2} is the only 
  one among these species present in the cloud. The spacing between 
  hash-marks symbolizes (but not to scale) changes in gas density.}
\end{figure}

\clearpage
\epsscale{1.0}
\begin{figure}
%\plotone{normcnkn.ps}
\plotone{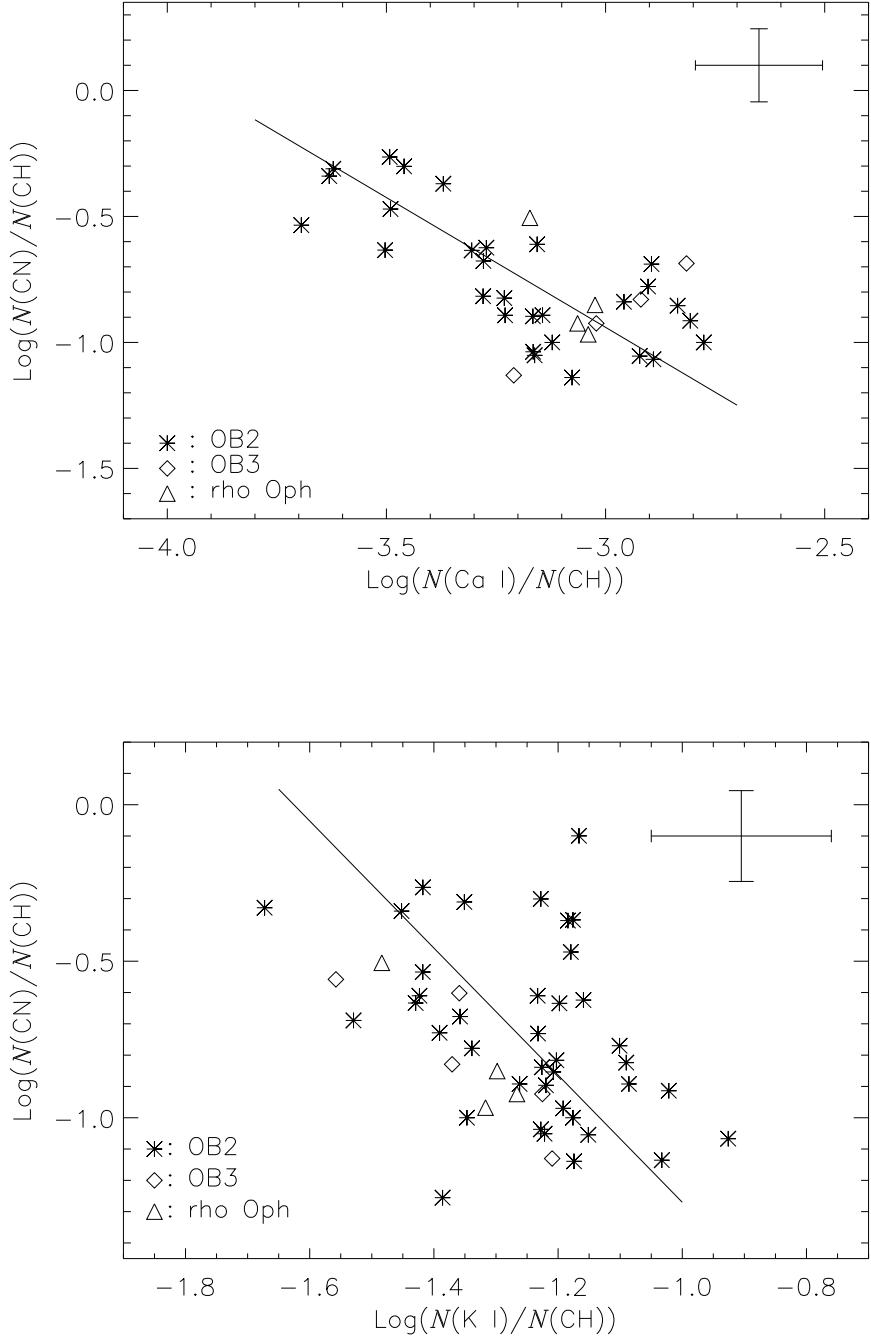}
\caption{Logarithmic plots of column density ratios. The error bars represent
20\% uncertainties in $N$.~~$Top$: $N$(CN)/$N$(CH) vs. 
$N$(\ion{Ca}{1})/$N$(CH). The solid line presents the best fit with a slope 
of $-$1.03 $\pm$ 0.15 ~~$Bottom$: $N$(CN)/$N$(CH) vs. 
$N$(\ion{K}{1})/$N$(CH). The solid line presents the best fit with a slope 
of $-$2.03 $\pm$ 0.20.}
\end{figure}

\clearpage

\begin{figure}
\epsscale{0.9}
\vspace{-0.50in}
%\plotone{cepob2_map.eps}
\plotone{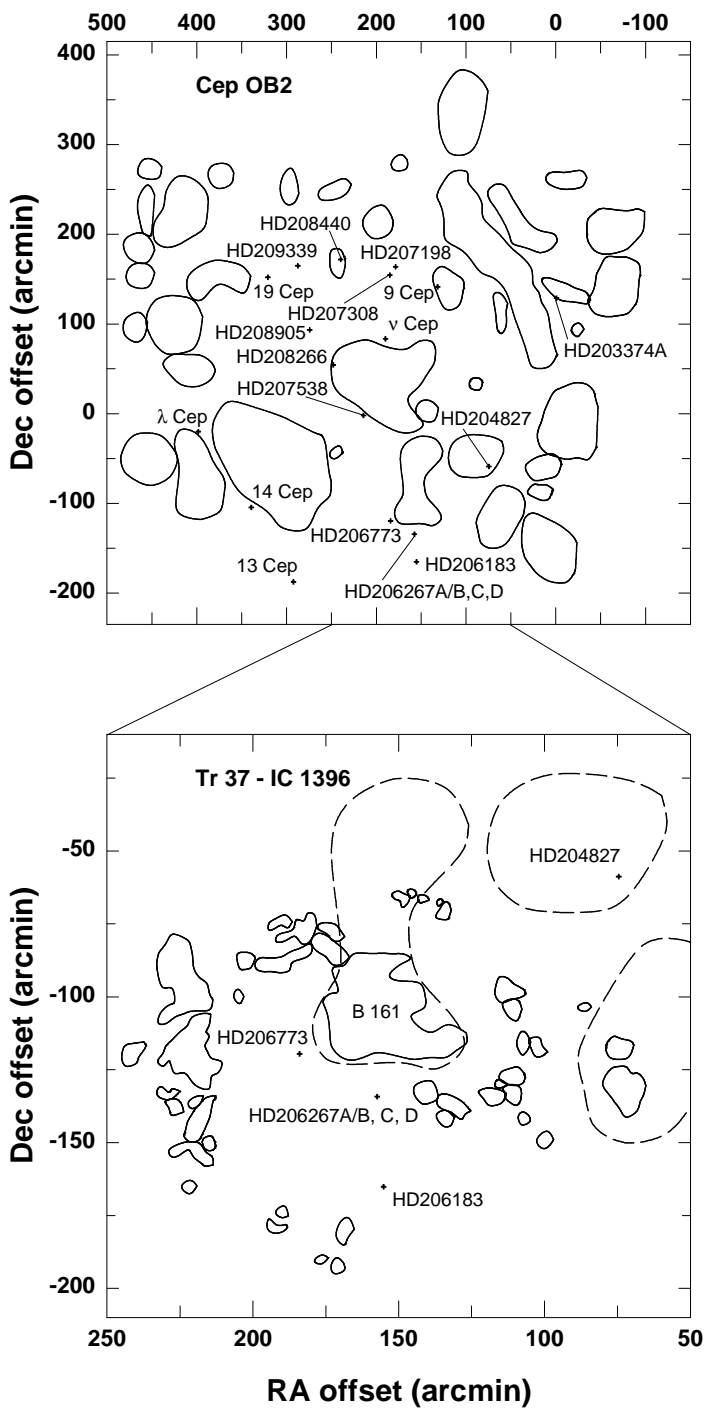}
\caption{Map of CO emission across the Cep OB2 Association, adopted from 
Patel et al. (1995, 1998). Stars observed in the present study are projected 
onto the map. The lower panel is an expanded view of Tr~37/IC~1396. The dashed 
contours represent the entities in the upper panel.}
\end{figure}

\clearpage

\begin{figure}
\epsscale{1.00}
\vspace{-0.20in}
%\plotone{h2chn.ps}
\plotone{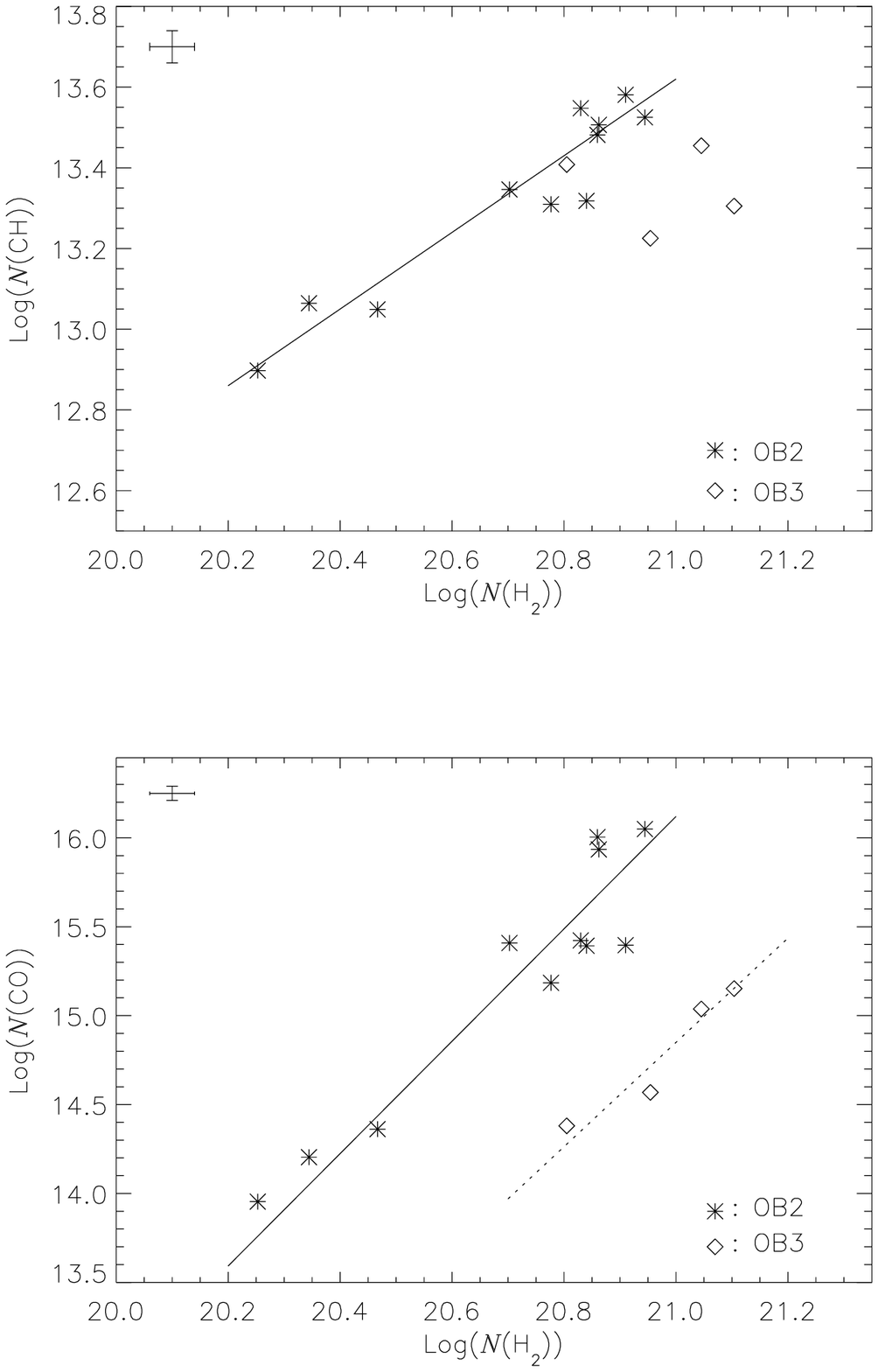}
\vspace{-0.80in}
\caption{Logarithmic plots of total column densities along lines of sight in
Cep OB2 and Cep OB3.  The error bars represent 10\% uncertainties in $N$. 
~~$Top$: $N$(CH) vs. $N$(H$_2$). The solid line shows the best fit to Cep OB2 data. 
~~$Bottom$: $N$(CO) vs. $N$(H$_2$). The solid line represents the best fit to
Cep OB2 data, and the dashed one to Cep OB3 data. Note that, for a given 
$N$(H$_2$), Cep OB2 sight lines have higher CH and CO column densities 
than lines of sight toward Cep OB3.}
\end{figure}
\end{document}